\documentclass[longauth]{aa}

\usepackage{graphicx}
\usepackage{txfonts}
\usepackage[normalem]{ulem}

\usepackage{natbib, twoopt}
\bibpunct{(}{)}{;}{a}{}{,}
\usepackage{orcidlink}

\makeatletter
\renewcommand*\maketitle{%
  \thispagestyle{firstpage}
\begingroup
    \if@wideboxfn
    \setlength\bibindent{1.4\parindent}
    \else
    \setlength\bibindent{\parindent}
    \fi
    \renewcommand*\thefootnote{\@fnsymbol\c@footnote}%
    \renewcommand\@makefntext[1]{%
    \ifaa@longfn\hsize\textwidth\fi
    \noindent
    \hb@xt@\bibindent{\hss\@makefnmark\enspace}##1}
  \ifaa@twocolumn
  \begingroup
  \begin{aa@strip}   
          \aa@maketitle
    \end{aa@strip}
    \@thanks
  \endgroup
  \else
    \begingroup
      \let\thanks\footnote
      \aa@maketitle
    \endgroup
  \fi
\endgroup
  \setcounter{footnote}{0}%
}
\makeatother

\begin{document}

\title{Constraining gamma-ray burst parameters with the first ultra-high energy neutrino event KM3-230213A}

\titlerunning{A\&A, XXX, A4 (2025)}
\authorrunning{The KM3NeT Collaboration}

\date{Received September XX, 2025}

\author{ 
O.~Adriani\,\orcidlink{0000-0002-3592-0654}\inst{\ref{b},\ref{a}} 
 \and A.~Albert\inst{\ref{c},\ref{bd}} 
 \and A.\,R.~Alhebsi\,\orcidlink{0009-0002-7320-7638}\inst{\ref{d}} 
 \and S.~Alshalloudi\inst{\ref{d}} 
 \and M.~Alshamsi\inst{\ref{e}} 
 \and S. Alves Garre\,\orcidlink{0000-0003-1893-0858}\inst{\ref{f}} 
 \and A. Ambrosone\inst{\ref{h},\ref{g}} 
 \and F.~Ameli\inst{\ref{i}} 
 \and M.~Andre\inst{\ref{j}} 
 \and L.~Aphecetche\,\orcidlink{0000-0001-7662-3878}\inst{\ref{k}} 
 \and M. Ardid\,\orcidlink{0000-0002-3199-594X}\inst{\ref{l}} 
 \and S. Ardid\,\orcidlink{0000-0003-4821-6655}\inst{\ref{l}} 
 \and J.~Aublin\inst{\ref{m}} 
 \and F.~Badaracco\,\orcidlink{0000-0001-8553-7904}\inst{\ref{o},\ref{n}} 
 \and L.~Bailly-Salins\inst{\ref{p}} 
 \and B.~Baret\inst{\ref{m}} 
 \and A. Bariego-Quintana\,\orcidlink{0000-0001-5187-7505}\inst{\ref{f}} 
 \and Y.~Becherini\inst{\ref{m}} 
 \and M.~Bendahman\inst{\ref{g}} 
 \and F.~Benfenati~Gualandi\inst{\ref{r},\ref{q}} 
 \and M.~Benhassi\inst{\ref{s},\ref{g}} 
 \and D.\,M.~Benoit\,\orcidlink{0000-0002-7773-6863}\inst{\ref{t}} 
 \and Z.~Be\v{n}u\v{s}ov\'a\,\orcidlink{0000-0002-2677-7657}\inst{\ref{v},\ref{u}} 
 \and E.~Berbee\inst{\ref{w}} 
 \and E.~Berti\inst{\ref{b}} 
 \and V.~Bertin\,\orcidlink{0000-0001-6688-4580}\inst{\ref{e}} 
 \and P.~Betti\inst{\ref{b}} 
 \and S.~Biagi\,\orcidlink{0000-0001-8598-0017}\inst{\ref{x}} 
 \and M.~Boettcher\inst{\ref{y}} 
 \and D.~Bonanno\,\orcidlink{0000-0003-0223-3580}\inst{\ref{x}} 
 \and S.~Bottai\inst{\ref{b}} 
 \and A.\,B.~Bouasla\inst{\ref{be}} 
 \and J.~Boumaaza\inst{\ref{z}} 
 \and M.~Bouta\inst{\ref{e}} 
 \and M.~Bouwhuis\inst{\ref{w}} 
 \and C.~Bozza\,\orcidlink{0000-0002-1797-6451}\inst{\ref{aa},\ref{g}} 
 \and R.\,M.~Bozza\inst{\ref{h},\ref{g}} 
 \and H.~Br\^anza\c s\inst{\ref{ab}} 
 \and F.~Bretaudeau\inst{\ref{k}} 
 \and M.~Breuhaus\,\orcidlink{0000-0003-0268-5122}\inst{\ref{e}} 
 \and R.~Bruijn\inst{\ref{ac},\ref{w}} 
 \and J.~Brunner\inst{\ref{e}} 
 \and R.~Bruno\,\orcidlink{0000-0002-3517-6597}\inst{\ref{ad}} 
 \and E.~Buis\inst{\ref{ae},\ref{w}} 
 \and R.~Buompane\inst{\ref{s},\ref{g}} 
 \and J.~Busto\inst{\ref{e}} 
 \and B.~Caiffi\inst{\ref{o}} 
 \and D.~Calvo\inst{\ref{f}} 
 \and A.~Capone\inst{\ref{i},\ref{af}} 
 \and F.~Carenini\inst{\ref{r},\ref{q}} 
 \and V.~Carretero\,\orcidlink{0000-0002-7540-0266}\inst{\ref{ac},\ref{w}} 
 \and T.~Cartraud\inst{\ref{m}} 
 \and P.~Castaldi\inst{\ref{ag},\ref{q}} 
 \and V.~Cecchini\,\orcidlink{0000-0003-4497-2584}\inst{\ref{f}} 
 \and S.~Celli\inst{\ref{i},\ref{af}} 
 \and L.~Cerisy\inst{\ref{e}} 
 \and M.~Chabab\inst{\ref{ah}} 
 \and A.~Chen\,\orcidlink{0000-0001-6425-5692}\inst{\ref{ai}} 
 \and S.~Cherubini\inst{\ref{aj},\ref{x}} 
 \and T.~Chiarusi\inst{\ref{q}} 
 \and M.~Circella\,\orcidlink{0000-0002-5560-0762}\inst{\ref{ak}} 
 \and R.~Clark\inst{\ref{al}} 
 \and R.~Cocimano\inst{\ref{x}} 
 \and J.\,A.\,B.~Coelho\inst{\ref{m}} 
 \and A.~Coleiro\inst{\ref{m}} 
 \and A. Condorelli\inst{\ref{m}} 
 \and R.~Coniglione\inst{\ref{x}} 
 \and P.~Coyle\inst{\ref{e}} 
 \and A.~Creusot\inst{\ref{m}} 
 \and G.~Cuttone\inst{\ref{x}} 
 \and R.~Dallier\,\orcidlink{0000-0001-9452-4849}\inst{\ref{k}} 
 \and A.~De~Benedittis\inst{\ref{s},\ref{g}} 
 \and G.~De~Wasseige\inst{\ref{al}}
 \and V.~Decoene\inst{\ref{k}} 
 \and P. Deguire\inst{\ref{e}} 
 \and I.~Del~Rosso\inst{\ref{r},\ref{q}} 
 \and L.\,S.~Di~Mauro\inst{\ref{x}} 
 \and I.~Di~Palma\inst{\ref{i},\ref{af}} 
 \and A.\,F.~D\'iaz\inst{\ref{am}} 
 \and D.~Diego-Tortosa\,\orcidlink{0000-0001-5546-3748}\inst{\ref{x}} 
 \and C.~Distefano\,\orcidlink{0000-0001-8632-1136}\inst{\ref{x}} 
 \and A.~Domi\inst{\ref{an}} 
 \and C.~Donzaud\inst{\ref{m}} 
 \and D.~Dornic\,\orcidlink{0000-0001-5729-1468}\inst{\ref{e}} 
 \and E.~Drakopoulou\,\orcidlink{0000-0003-2493-8039}\inst{\ref{ao}} 
 \and D.~Drouhin\,\orcidlink{0000-0002-9719-2277}\inst{\ref{c},\ref{bd}} 
 \and J.-G. Ducoin\inst{\ref{e}} 
 \and P.~Duverne\inst{\ref{m}} 
 \and R. Dvornick\'{y}\,\orcidlink{0000-0002-4401-1188}\inst{\ref{v}} 
 \and T.~Eberl\,\orcidlink{0000-0002-5301-9106}\inst{\ref{an}} 
 \and E. Eckerov\'{a}\inst{\ref{v},\ref{u}} 
 \and A.~Eddymaoui\inst{\ref{z}} 
 \and T.~van~Eeden\inst{\ref{w}} 
 \and M.~Eff\inst{\ref{m}} 
 \and D.~van~Eijk\inst{\ref{w}} 
 \and I.~El~Bojaddaini\inst{\ref{ap}} 
 \and S.~El~Hedri\inst{\ref{m}} 
 \and S.~El~Mentawi\inst{\ref{e}} 
 \and A.~Enzenh\"ofer\inst{\ref{e}} 
 \and G.~Ferrara\inst{\ref{aj},\ref{x}} 
 \and M.~D.~Filipovi\'c\,\orcidlink{0000-0002-4990-9288}\inst{\ref{aq}} 
 \and F.~Filippini\inst{\ref{q}} 
 \and D.~Franciotti\inst{\ref{x}} 
 \and L.\,A.~Fusco\inst{\ref{aa},\ref{g}} 
 \and S.~Gagliardini\inst{\ref{af},\ref{i}} 
 \and T.~Gal\,\orcidlink{0000-0001-7821-8673}\inst{\ref{an}} 
 \and J.~Garc{\'\i}a~M{\'e}ndez\,\orcidlink{0000-0002-1580-0647}\inst{\ref{l}} 
 \and A.~Garcia~Soto\,\orcidlink{0000-0002-8186-2459}\inst{\ref{f}} 
 \and C.~Gatius~Oliver\,\orcidlink{0009-0002-1584-1788}\inst{\ref{w}} 
 \and N.~Gei{\ss}elbrecht\inst{\ref{an}} 
 \and E.~Genton\inst{\ref{al}} 
 \and H.~Ghaddari\inst{\ref{ap}} 
 \and L.~Gialanella\inst{\ref{s},\ref{g}} 
 \and B.\,K.~Gibson\inst{\ref{t}} 
 \and E.~Giorgio\inst{\ref{x}} 
 \and I.~Goos\,\orcidlink{0009-0008-1479-539X}\inst{\ref{m}} 
 \and P.~Goswami\inst{\ref{m}} 
 \and S.\,R.~Gozzini\,\orcidlink{0000-0001-5152-9631}\inst{\ref{f}} 
 \and R.~Gracia\inst{\ref{an}} 
 \and B.~Guillon\inst{\ref{p}} 
 \and C.~Haack\inst{\ref{an}} 
 \and H.~van~Haren\inst{\ref{ar}} 
 \and A.~Heijboer\inst{\ref{w}} 
 \and L.~Hennig\inst{\ref{an}} 
 \and J.\,J.~Hern{\'a}ndez-Rey\inst{\ref{f}} 
 \and A.~Idrissi\,\orcidlink{0000-0001-8936-6364}\inst{\ref{x}} 
 \and W.~Idrissi~Ibnsalih\inst{\ref{g}} 
 \and G.~Illuminati\inst{\ref{q}} 
 \and R.~Jaimes\inst{\ref{f}} 
 \and O.~Janik\inst{\ref{an}} 
 \and D.~Joly\inst{\ref{e}} 
 \and M.~de~Jong\inst{\ref{as},\ref{w}} 
 \and P.~de~Jong\inst{\ref{ac},\ref{w}} 
 \and B.\,J.~Jung\inst{\ref{w}} 
 \and P.~Kalaczy\'nski\,\orcidlink{0000-0001-9278-5906}\inst{\ref{bf},\ref{at}} 
 \and J.~Keegans\inst{\ref{t}} 
 \and V.~Kikvadze\inst{\ref{au}} 
 \and G.~Kistauri\inst{\ref{av},\ref{au}} 
 \and C.~Kopper\,\orcidlink{0000-0001-6288-7637}\inst{\ref{an}} 
 \and A.~Kouchner\inst{\ref{aw},\ref{m}} 
 \and Y. Y. Kovalev\,\orcidlink{0000-0001-9303-3263}\inst{\ref{ax}} 
 \and L.~Krupa\inst{\ref{u}} 
 \and V.~Kueviakoe\inst{\ref{w}} 
 \and V.~Kulikovskiy\inst{\ref{o}} 
 \and R.~Kvatadze\inst{\ref{av}} 
 \and M.~Labalme\inst{\ref{p}} 
 \and R.~Lahmann\inst{\ref{an}} 
 \and M.~Lamoureux\,\orcidlink{0000-0002-8860-5826}\inst{\ref{al}} 
 \and G.~Larosa\inst{\ref{x}} 
 \and C.~Lastoria\inst{\ref{p}} 
 \and J.~Lazar\inst{\ref{al}} 
 \and A.~Lazo\inst{\ref{f}} 
 \and G.~Lehaut\inst{\ref{p}} 
 \and V.~Lema\^itre\inst{\ref{al}} 
 \and E.~Leonora\inst{\ref{ad}} 
 \and N.~Lessing\inst{\ref{f}} 
 \and G.~Levi\inst{\ref{r},\ref{q}} 
 \and M.~Lindsey~Clark\inst{\ref{m}} 
 \and F.~Longhitano\inst{\ref{ad}} 
 \and S.~Madarapu\inst{\ref{f}} 
 \and F.~Magnani\inst{\ref{e}} 
 \and L.~Malerba\inst{\ref{o},\ref{n}} 
 \and F.~Mamedov\inst{\ref{u}} 
 \and A.~Manfreda\,\orcidlink{0000-0002-0998-4953}\inst{\ref{g}} 
 \and A.~Manousakis\inst{\ref{ay}} 
 \and M.~Marconi\,\orcidlink{0009-0008-0023-4647}\inst{\ref{n},\ref{o}} 
 \and A.~Margiotta\,\orcidlink{0000-0001-6929-5386}\inst{\ref{r},\ref{q}} 
 \and A.~Marinelli\inst{\ref{h},\ref{g}} 
 \and C.~Markou\inst{\ref{ao}} 
 \and L.~Martin\,\orcidlink{0000-0002-9781-2632}\inst{\ref{k}} 
 \and M.~Mastrodicasa\inst{\ref{af},\ref{i}} 
 \and S.~Mastroianni\inst{\ref{g}} 
 \and J.~Mauro\,\orcidlink{0009-0005-9324-7970}\inst{\ref{al}} 
 \and K.\,C.\,K.~Mehta\inst{\ref{at}} 
 \and G.~Miele\inst{\ref{h},\ref{g}} 
 \and P.~Migliozzi\,\orcidlink{0000-0001-5497-3594}\inst{\ref{g}} 
 \and E.~Migneco\inst{\ref{x}} 
 \and M.\,L.~Mitsou\inst{\ref{s},\ref{g}} 
 \and C.\,M.~Mollo\inst{\ref{g}} 
 \and L. Morales-Gallegos\,\orcidlink{0000-0002-2241-4365}\inst{\ref{s},\ref{g}} 
 \and N.~Mori\,\orcidlink{0000-0003-2138-3787}\inst{\ref{b}} 
 \and A.~Moussa\,\orcidlink{0000-0003-2233-9120}\inst{\ref{ap}} 
 \and I.~Mozun~Mateo\inst{\ref{p}} 
 \and R.~Muller\,\orcidlink{0000-0002-5247-7084}\inst{\ref{q}} 
 \and M.\,R.~Musone\inst{\ref{s},\ref{g}} 
 \and M.~Musumeci\inst{\ref{x}} 
 \and S.~Navas\,\orcidlink{0000-0003-1688-5758}\inst{\ref{az}} 
 \and A.~Nayerhoda\inst{\ref{ak}} 
 \and C.\,A.~Nicolau\inst{\ref{i}} 
 \and B.~Nkosi\,\orcidlink{0000-0003-0954-4779}\inst{\ref{ai}} 
 \and B.~{\'O}~Fearraigh\,\orcidlink{0000-0002-1795-1617}\inst{\ref{o}} 
 \and V.~Oliviero\,\orcidlink{0009-0004-9638-0825}\inst{\ref{h},\ref{g}} 
 \and A.~Orlando\inst{\ref{x}} 
 \and E.~Oukacha\inst{\ref{m}} 
 \and L.~Pacini\,\orcidlink{0000-0001-6808-9396}\inst{\ref{b}} 
 \and D.~Paesani\inst{\ref{x}} 
 \and J.~Palacios~Gonz{\'a}lez\,\orcidlink{0000-0001-9292-9981}\inst{\ref{f}} 
 \and G.~Papalashvili\inst{\ref{ak},\ref{au}} 
 \and P.~Papini\inst{\ref{b}} 
 \and V.~Parisi\inst{\ref{n},\ref{o}} 
 \and A.~Parmar\inst{\ref{p}} 
 \and C.~Pastore\inst{\ref{ak}} 
 \and A.~M.~P\u{a}un\inst{\ref{ab}} 
 \and G.\,E.~P\u{a}v\u{a}la\c{s}\inst{\ref{ab}} 
 \and S. Pe\~{n}a Mart\'inez\,\orcidlink{0000-0001-8939-0639}\inst{\ref{m}} 
 \and M.~Perrin-Terrin\inst{\ref{e}} 
 \and V.~Pestel\inst{\ref{p}} 
 \and M.~Petropavlova\,\orcidlink{0000-0002-0416-0795}\inst{\ref{u},\ref{bg}} 
 \and P.~Piattelli\inst{\ref{x}} 
 \and A.~Plavin\inst{\ref{ax},\ref{bh}} 
 \and C.~Poir{\`e}\inst{\ref{aa},\ref{g}} 
 \and V.~Popa\inst{\ref{ab}} \thanks{Deceased}
 \and T.~Pradier\,\orcidlink{0000-0001-5501-0060}\inst{\ref{c}} 
 \and J.~Prado\inst{\ref{f}} 
 \and S.~Pulvirenti\inst{\ref{x}} 
 \and C.A.~Quiroz-Rangel\,\orcidlink{0009-0002-3446-8747}\inst{\ref{l}} 
 \and N.~Randazzo\inst{\ref{ad}} 
 \and A.~Ratnani\inst{\ref{ba}} 
 \and S.~Razzaque\,\orcidlink{0000-0002-0130-2460}\inst{\ref{bb}}\thanks{Corresponding authors: \email{km3net-pc@km3net.de}, psevle@km3net.de, srazzaque@km3net.de}
 \and I.\,C.~Rea\,\orcidlink{0000-0002-3954-7754}\inst{\ref{g}} 
 \and D.~Real\,\orcidlink{0000-0002-1038-7021}\inst{\ref{f}} 
 \and G.~Riccobene\,\orcidlink{0000-0002-0600-2774}\inst{\ref{x}} 
 \and J.~Robinson\inst{\ref{y}} 
 \and A.~Romanov\inst{\ref{n},\ref{o},\ref{p}} 
 \and E.~Ros\inst{\ref{ax}} 
 \and A. \v{S}aina\inst{\ref{f}} 
 \and F.~Salesa~Greus\,\orcidlink{0000-0002-8610-8703}\inst{\ref{f}} 
 \and D.\,F.\,E.~Samtleben\inst{\ref{as},\ref{w}} 
 \and A.~S{\'a}nchez~Losa\,\orcidlink{0000-0001-9596-7078}\inst{\ref{f}} 
 \and S.~Sanfilippo\inst{\ref{x}} 
 \and M.~Sanguineti\inst{\ref{n},\ref{o}} 
 \and D.~Santonocito\inst{\ref{x}} 
 \and P.~Sapienza\inst{\ref{x}} 
 \and M.~Scaringella\inst{\ref{b}} 
 \and M.~Scarnera\inst{\ref{al},\ref{m}} 
 \and J.~Schnabel\inst{\ref{an}} 
 \and J.~Schumann\,\orcidlink{0000-0003-3722-086X}\inst{\ref{an}} 
 \and J.~Seneca\inst{\ref{w}} 
 \and P. A.~Sevle~Myhr\,\orcidlink{0009-0005-9103-4410}\inst{\ref{al}}$^{\star\star}$
 \and I.~Sgura\inst{\ref{ak}} 
 \and R.~Shanidze\inst{\ref{au}} 
 \and Chengyu Shao\,\orcidlink{0000-0002-2954-1180}\inst{\ref{bi},\ref{e}} 
 \and A.~Sharma\inst{\ref{m}} 
 \and Y.~Shitov\inst{\ref{u}} 
 \and F. \v{S}imkovic\inst{\ref{v}} 
 \and A.~Simonelli\inst{\ref{g}} 
 \and A.~Sinopoulou\,\orcidlink{0000-0001-9205-8813}\inst{\ref{ad}} 
 \and B.~Spisso\inst{\ref{g}} 
 \and M.~Spurio\,\orcidlink{0000-0002-8698-3655}\inst{\ref{r},\ref{q}} 
 \and O.~Starodubtsev\inst{\ref{b}} 
 \and D.~Stavropoulos\inst{\ref{ao}} 
 \and I. \v{S}tekl\inst{\ref{u}} 
 \and D.~Stocco\,\orcidlink{0000-0002-5377-5163}\inst{\ref{k}} 
 \and M.~Taiuti\inst{\ref{n},\ref{o}} 
 \and G.~Takadze\inst{\ref{au}} 
 \and Y.~Tayalati\inst{\ref{z},\ref{ba}} 
 \and H.~Thiersen\inst{\ref{y}} 
 \and S.~Thoudam\inst{\ref{d}} 
 \and I.~Tosta~e~Melo\inst{\ref{ad},\ref{aj}} 
 \and B.~Trocm{\'e}\,\orcidlink{0000-0001-9500-2487}\inst{\ref{m}} 
 \and V.~Tsourapis\,\orcidlink{0009-0000-5616-5662}\inst{\ref{ao}} 
 \and E.~Tzamariudaki\inst{\ref{ao}} 
 \and A.~Ukleja\,\orcidlink{0000-0003-0480-4850}\inst{\ref{at}} 
 \and A.~Vacheret\inst{\ref{p}} 
 \and V.~Valsecchi\inst{\ref{x}} 
 \and V.~Van~Elewyck\inst{\ref{aw},\ref{m}} 
 \and G.~Vannoye\inst{\ref{n},\ref{o}} 
 \and E.~Vannuccini\inst{\ref{b}} 
 \and G.~Vasileiadis\inst{\ref{bc}} 
 \and F.~Vazquez~de~Sola\inst{\ref{w}} 
 \and A. Veutro\inst{\ref{i},\ref{af}} 
 \and S.~Viola\inst{\ref{x}} 
 \and D.~Vivolo\inst{\ref{s},\ref{g}} 
 \and A. van Vliet\,\orcidlink{0000-0003-2827-3361}\inst{\ref{d}} 
 \and E.~de~Wolf\,\orcidlink{0000-0002-8272-8681}\inst{\ref{ac},\ref{w}} 
 \and I.~Lhenry-Yvon\inst{\ref{m}} 
 \and S.~Zavatarelli\inst{\ref{o}} 
 \and D.~Zito\inst{\ref{x}} 
 \and J.\,D.~Zornoza\,\orcidlink{0000-0002-1834-0690}\inst{\ref{f}} 
 \and J.~Z{\'u}{\~n}iga\,\orcidlink{0000-0002-1041-6451}\inst{\ref{f}} 
}

\institute{ 
Universit{\`a} di Firenze, Dipartimento di Fisica e Astronomia, via Sansone 1, Sesto Fiorentino, 50019 Italy\label{a} 
 \and INFN, Sezione di Firenze, via Sansone 1, Sesto Fiorentino, 50019 Italy\label{b} 
 \and Universit{\'e}~de~Strasbourg,~CNRS,~IPHC~UMR~7178,~F-67000~Strasbourg,~France\label{c} 
 \and Khalifa University of Science and Technology, Department of Physics, PO Box 127788, Abu Dhabi,   United Arab Emirates\label{d} 
 \and Aix~Marseille~Univ,~CNRS/IN2P3,~CPPM,~Marseille,~France\label{e} 
 \and IFIC - Instituto de F{\'\i}sica Corpuscular (CSIC - Universitat de Val{\`e}ncia), c/Catedr{\'a}tico Jos{\'e} Beltr{\'a}n, 2, 46980 Paterna, Valencia, Spain\label{f} 
 \and INFN, Sezione di Napoli, Complesso Universitario di Monte S. Angelo, Via Cintia ed. G, Napoli, 80126 Italy\label{g} 
 \and Universit{\`a} di Napoli ``Federico II'', Dip. Scienze Fisiche ``E. Pancini'', Complesso Universitario di Monte S. Angelo, Via Cintia ed. G, Napoli, 80126 Italy\label{h} 
 \and INFN, Sezione di Roma, Piazzale Aldo Moro, 2 - c/o Dipartimento di Fisica, Edificio, G.Marconi, Roma, 00185 Italy\label{i} 
 \and Universitat Polit{\`e}cnica de Catalunya, Laboratori d'Aplicacions Bioac{\'u}stiques, Centre Tecnol{\`o}gic de Vilanova i la Geltr{\'u}, Avda. Rambla Exposici{\'o}, s/n, Vilanova i la Geltr{\'u}, 08800 Spain\label{j} 
 \and Subatech, IMT Atlantique, IN2P3-CNRS, Nantes Universit{\'e}, 4 rue Alfred Kastler - La Chantrerie, Nantes, BP 20722 44307 France\label{k} 
 \and Universitat Polit{\`e}cnica de Val{\`e}ncia, Instituto de Investigaci{\'o}n para la Gesti{\'o}n Integrada de las Zonas Costeras, C/ Paranimf, 1, Gandia, 46730 Spain\label{l} 
 \and Universit{\'e} Paris Cit{\'e}, CNRS, Astroparticule et Cosmologie, F-75013 Paris, France\label{m} 
 \and Universit{\`a} di Genova, Via Dodecaneso 33, Genova, 16146 Italy\label{n} 
 \and INFN, Sezione di Genova, Via Dodecaneso 33, Genova, 16146 Italy\label{o} 
 \and LPC CAEN, Normandie Univ, ENSICAEN, UNICAEN, CNRS/IN2P3, 6 boulevard Mar{\'e}chal Juin, Caen, 14050 France\label{p} 
 \and INFN, Sezione di Bologna, v.le C. Berti-Pichat, 6/2, Bologna, 40127 Italy\label{q} 
 \and Universit{\`a} di Bologna, Dipartimento di Fisica e Astronomia, v.le C. Berti-Pichat, 6/2, Bologna, 40127 Italy\label{r} 
 \and Universit{\`a} degli Studi della Campania "Luigi Vanvitelli", Dipartimento di Matematica e Fisica, viale Lincoln 5, Caserta, 81100 Italy\label{s} 
 \and E.\,A.~Milne Centre for Astrophysics, University~of~Hull, Hull, HU6 7RX, United Kingdom\label{t} 
 \and Czech Technical University in Prague, Institute of Experimental and Applied Physics, Husova 240/5, Prague, 110 00 Czech Republic\label{u} 
 \and Comenius University in Bratislava, Department of Nuclear Physics and Biophysics, Mlynska dolina F1, Bratislava, 842 48 Slovak Republic\label{v} 
 \and Nikhef, National Institute for Subatomic Physics, PO Box 41882, Amsterdam, 1009 DB Netherlands\label{w} 
 \and INFN, Laboratori Nazionali del Sud, (LNS) Via S. Sofia 62, Catania, 95123 Italy\label{x} 
 \and North-West University, Centre for Space Research, Private Bag X6001, Potchefstroom, 2520 South Africa\label{y} 
 \and University Mohammed V in Rabat, Faculty of Sciences, 4 av.~Ibn Battouta, B.P.~1014, R.P.~10000 Rabat, Morocco\label{z} 
 \and Universit{\`a} di Salerno e INFN Gruppo Collegato di Salerno, Dipartimento di Fisica, Via Giovanni Paolo II 132, Fisciano, 84084 Italy\label{aa} 
 \and Institute of Space Science - INFLPR Subsidiary, 409 Atomistilor Street, Magurele, Ilfov, 077125 Romania\label{ab} 
 \and University of Amsterdam, Institute of Physics/IHEF, PO Box 94216, Amsterdam, 1090 GE Netherlands\label{ac} 
 \and INFN, Sezione di Catania, (INFN-CT) Via Santa Sofia 64, Catania, 95123 Italy\label{ad} 
 \and TNO, Technical Sciences, PO Box 155, Delft, 2600 AD Netherlands\label{ae} 
 \and Universit{\`a} La Sapienza, Dipartimento di Fisica, Piazzale Aldo Moro 2, Roma, 00185 Italy\label{af} 
 \and Universit{\`a} di Bologna, Dipartimento di Ingegneria dell'Energia Elettrica e dell'Informazione "Guglielmo Marconi", Via dell'Universit{\`a} 50, Cesena, 47521 Italia\label{ag} 
 \and Cadi Ayyad University, Physics Department, Faculty of Science Semlalia, Av. My Abdellah, P.O.B. 2390, Marrakech, 40000 Morocco\label{ah} 
 \and University of the Witwatersrand, School of Physics, Private Bag 3, Johannesburg, Wits 2050 South Africa\label{ai} 
 \and Universit{\`a} di Catania, Dipartimento di Fisica e Astronomia "Ettore Majorana", (INFN-CT) Via Santa Sofia 64, Catania, 95123 Italy\label{aj} 
 \and INFN, Sezione di Bari, via Orabona, 4, Bari, 70125 Italy\label{ak} 
 \and UCLouvain, Centre for Cosmology, Particle Physics and Phenomenology, Chemin du Cyclotron, 2, Louvain-la-Neuve, 1348 Belgium\label{al} 
 \and University of Granada, Department of Computer Engineering, Automation and Robotics / CITIC, 18071 Granada, Spain\label{am} 
 \and Friedrich-Alexander-Universit{\"a}t Erlangen-N{\"u}rnberg (FAU), Erlangen Centre for Astroparticle Physics, Nikolaus-Fiebiger-Stra{\ss}e 2, 91058 Erlangen, Germany\label{an} 
 \and NCSR Demokritos, Institute of Nuclear and Particle Physics, Ag. Paraskevi Attikis, Athens, 15310 Greece\label{ao} 
 \and University Mohammed I, Faculty of Sciences, BV Mohammed VI, B.P.~717, R.P.~60000 Oujda, Morocco\label{ap} 
 \and Western Sydney University, School of Science, Locked Bag 1797, Penrith, NSW 2751 Australia\label{aq} 
 \and NIOZ (Royal Netherlands Institute for Sea Research), PO Box 59, Den Burg, Texel, 1790 AB, the Netherlands\label{ar} 
 \and Leiden University, Leiden Institute of Physics, PO Box 9504, Leiden, 2300 RA Netherlands\label{as} 
 \and AGH University of Krakow, Al.~Mickiewicza 30, 30-059 Krakow, Poland\label{at} 
 \and Tbilisi State University, Department of Physics, 3, Chavchavadze Ave., Tbilisi, 0179 Georgia\label{au} 
 \and The University of Georgia, Institute of Physics, Kostava str. 77, Tbilisi, 0171 Georgia\label{av} 
 \and Institut Universitaire de France, 1 rue Descartes, Paris, 75005 France\label{aw} 
 \and Max-Planck-Institut~f{\"u}r~Radioastronomie,~Auf~dem H{\"u}gel~69,~53121~Bonn,~Germany\label{ax} 
 \and University of Sharjah, Sharjah Academy for Astronomy, Space Sciences, and Technology, University Campus - POB 27272, Sharjah, - United Arab Emirates\label{ay} 
 \and University of Granada, Dpto.~de F\'\i{}sica Te\'orica y del Cosmos \& C.A.F.P.E., 18071 Granada, Spain\label{az} 
 \and School of Applied and Engineering Physics, Mohammed VI Polytechnic University, Ben Guerir, 43150, Morocco\label{ba} 
 \and University of Johannesburg, Department Physics, PO Box 524, Auckland Park, 2006 South Africa\label{bb} 
 \and Laboratoire Univers et Particules de Montpellier, Place Eug{\`e}ne Bataillon - CC 72, Montpellier C{\'e}dex 05, 34095 France\label{bc} 
 \and Universit{\'e} de Haute Alsace, rue des Fr{\`e}res Lumi{\`e}re, 68093 Mulhouse Cedex, France\label{bd} 
 \and Universit{\'e} Badji Mokhtar, D{\'e}partement de Physique, Facult{\'e} des Sciences, Laboratoire de Physique des Rayonnements, B. P. 12, Annaba, 23000 Algeria\label{be} 
 \and AstroCeNT, Nicolaus Copernicus Astronomical Center, Polish Academy of Sciences, Rektorska 4, Warsaw, 00-614 Poland\label{bf} 
 \and Charles University, Faculty of Mathematics and Physics, Ovocn{\'y} trh 5, Prague, 116 36 Czech Republic\label{bg} 
 \and Harvard University, Black Hole Initiative, 20 Garden Street, Cambridge, MA 02138 USA\label{bh} 
 \and School~of~Physics~and~Astronomy, Sun Yat-sen University, Zhuhai, China\label{bi} 
}

\abstract
{The detection of the highest energy neutrino observed to date by KM3NeT, with an estimated energy of 220 PeV, opens up new possibilities for the study and identification of the astrophysical sources responsible for a diffuse flux of such ultra-high-energy neutrinos, among which gamma-ray bursts are longstanding candidates.}
{Based on the event KM3-230213A, we derived constraints on the baryon loading and density of the surrounding environment in models of blast waves in long-duration gamma-ray bursts.}
{We computed the diffuse flux from gamma-ray burst blast waves, either expanding in a constant density interstellar medium or developing in a radially decreasing density of a wind-like environment surrounding the gamma-ray burst progenitor star, by taking into account the expected neutrino spectra and luminosity function. We used a Poisson likelihood method to constrain the blast wave model parameters by calculating the expected number of neutrino events within the 90\% confidence level energy range of KM3-230213A and by using the joint exposure of KM3NeT/ARCA, IceCube, and Pierre Auger.}
{We constrain the baryon loading to be $f_b\leq 51$ at 90\% confidence, with the best-fit and 68\% confidence interval being $f_b=26.9^{+11.4}_{-17.2}$ for a constant interstellar medium particle density of $n_0=1$ cm$^{-3}$. In the wind-like environment case, the baryon loading is $f_b\leq 1095$ at 90\% confidence, with the corresponding 68\% confidence interval being $f_b\in[8,231]$, which is proportional to the sixth power of a variable density parameter of $A_*=0.1$.}
{}

\keywords{astroparticle physics -- neutrinos -- stars: gamma-ray bursts}

\maketitle

\section{Introduction}
The KM3NeT Collaboration reported the observation of a neutrino with an estimated $E_\nu=220^{+570}_{-110}$ PeV with the KM3NeT/ARCA detector in a partial configuration. It is the highest energy neutrino observed to date.
The event represents the first neutrino of presumable astrophysical origin observed in the ultra-high-energy (UHE) regime. Subsequent studies explored how this event fits into the global neutrino landscape, taking into account the lack of positive detections reported so far by the IceCube (IC;~\citet{Meier:2024flg}) and Pierre Auger (Auger;~\citet{PierreAuger:2023pjg}) Observatories. By combining these different (non-)observations, the most likely single-flavour diffuse astrophysical neutrino flux required to produce an event such as KM3-230213A is $E^2\Phi^{\rm 1f}_{\nu+\bar\nu}= 7.5^{+13.1}_{-4.7}\times 10^{-10}$ GeV~cm$^{-2}$~s$^{-1}$~sr$^{-1}$, assuming an $E^{-2}$ neutrino flux \citep{KM3NeT:2025global}.

Since its discovery in 2013 \citep{doi:10.1126/science.1242856}, the diffuse flux of astrophysical neutrinos between TeV and PeV energies has been extensively investigated \citep[see e.g.][]{Halzen2024arXiv241115329H}, and a multitude of potential sources have been scrutinised. Despite the identification of a few likely sources \citep{icecube2018neutrino, icecube2022evidence, icecube2023observation}, the origin of the majority of the diffuse neutrino flux remains unknown. The lack of observed neutrino multiplets \citep{abbasi2025search} further suggests that the population of neutrino-producing astrophysical objects consists of relatively dim, abundant, and isotropically distributed sources \citep{murase2016constraining}.

Gamma-ray bursts (GRBs) are the most energetic transient events observed in the Universe in electromagnetic wavebands and are potential sources of UHE cosmic rays ($E\gtrsim 10^{18}$~eV) \citep{Waxman1995PhRvL..75..386W, Vietri1995ApJ...453..883V}. Consequent neutrino production has been predicted from interactions of cosmic-ray protons with photons within the fireball in internal shocks  \citep{Waxman1997PhRvL..78.2292W}. High-energy neutrinos may also be created when the outgoing blast wave, in which particles are accelerated by 
external shocks, interacts with matter and radiation fields surrounding the GRB \citep{Waxman2000ApJ...541..707W, Dai2001ApJ...551..249D}. Previous searches for GRB neutrinos yielded no detection~\citep{IC_GRB_2016ApJ...824..115A, 10.1093/mnras/stx902, IC_GRB_2022ApJ...939..116A}. These searches have focused on analysing triggering GRBs -- GRBs bright enough in gamma rays to initiate multi-wavelength follow-ups -- and coincident neutrinos from varying time frames around the prompt emission phase of the GRBs. Although these searches have so far not resulted in direct detection, constraints have been put on the ratio of energy between protons and electrons, known as `baryon loading'~\citep{Rees1994ApJ...430L..93R}. This method, albeit successful in constraining individual GRB models, is not sensitive to the potentially large population of GRBs undetected by gamma-ray satellites \citep[see e.g.][]{Maggie2024arXiv241107973L}. Furthermore, the imposed constraint on the baryon loading depends on the GRB prompt emission model, which is uncertain, as well as on the model parameters. The purpose of this paper is therefore to investigate whether a larger population of GRBs can produce a significant fraction of the diffuse UHE neutrino flux, with emphasis on the undetected part of this population.

Previous constraints on the diffuse UHE neutrino flux have been set by both the IceCube and Pierre Auger observatories \citep{IceCube-Gen2:2020qha, PierreAuger:2019azx}. The lack of detection of UHE neutrinos by these observatories made the upper limits the strongest constraints available until the detection of KM3-230213A.

In this paper, we use the recent observation of a UHE neutrino event to constrain the baryon loading of GRB blast waves and the density of the medium in which the blast wave interacts to produce PeV--EeV neutrinos. We used the first-ever observation of a UHE neutrino to constrain the total contribution of
GRB blast waves to the diffuse UHE neutrino flux and consequently estimated some of the relevant model parameters. This paper is structured as follows: In Section 2 we outline the model under consideration for GRB blast wave neutrino production and how a large population of these GRBs can generate a diffuse extragalactic neutrino flux capable of producing KM3-230213A. In Section 3, we present an innovative technique to calculate the diffuse UHE neutrino flux, and the statistical framework used to put constraints on the GRB model parameters. In Section 4 we report and discusses the obtained constraints and their implications on GRBs. Finally, we summarise our findings in Section 5 and highlight our conclusions therein.

\section{Diffuse GRB neutrino flux in the PeV---EeV range}
We investigated the possible constraints enforced by KM3-230213A on specific model parameters of long-duration GRBs (lGRBs) by considering the contribution of a large population of lGRBs, up to redshift $z=5$, to the diffuse neutrino flux at UHEs. Despite a more prominent prompt emission phase at lower neutrino energies~\citep{Waxman1997PhRvL..78.2292W}, a significant amount of energy in lGRBs is expected to be converted to kinetic energy in the form of protons propagating outwards from the GRB in a blast wave that subsequently interacts with surrounding matter and radiation fields~\citep{Razzaque2013PhRvD..88j3003R}. The kinetic energy of the blast wave, $E_k$, is connected to the inferred gamma-ray luminosity during the prompt phase, $L_\gamma$, through the relation
\begin{equation}
    \label{eq:ratio}
    \frac{E_k}{L_\gamma}=f_b\eta t^*,
\end{equation}
where $f_b$ is the baryon loading ratio, $\eta$ is the efficiency of converting kinetic energy to gamma-ray energy, and $t^*$ is the timescale for the prompt emission. In our calculations, we set $\eta=0.2$ \citep{Fan2006MNRAS.369..197F} and adopted a normal distribution for $\log(t^*)$, fit to the distribution of lGRBs reported in \citet{vonKienlin:2020xvz}, as $\log(t^*)\sim\mathcal{N}(\mu,\sigma^2)$ with $\mu=1.44$ and $\sigma=0.50$.

\subsection{GRB blast wave models}
After the prompt emission phase, relativistic ejecta from the GRB drive a blast wave interacting in the gaseous media surrounding the GRB progenitor system. The blast wave expands and cools adiabatically, and after a deceleration time ($t_{\rm dec}$), its evolution is described by a simple similarity solution \citep{Blandford1976PhFl...19.1130B}. Synchrotron radiation by electrons in the surrounding gas, accelerating and cooling in the magnetic field in the shock region (forward shock), explains the radio-to-gamma-ray afterglow emission from GRB afterglows \citep{Meszaros1997ApJ...476..232M, Sari1998ApJ...497L..17S}.
Protons are co-accelerated with electrons to UHEs in the reverse shock, providing material to the ejecta and in the blast wave itself. These protons then interact with synchrotron photons to produce UHE neutrinos through photopion interactions \citep[see e.g.][]{Waxman1997PhRvL..78.2292W, Dai2001ApJ...551..249D, Murase2007PhRvD..76l3001M, Razzaque2013PhRvD..88j3003R}.
In the following, we consider two different scenarios: (1) an adiabatic blast wave interacting with the interstellar medium (ISM) of constant density and (2) a wind-type environment (WIND) around the GRB progenitor system with its density falling as $R^{-2}$, where $R$ is the distance from the GRB. Protons accelerated in the forward shock interact with afterglow synchrotron photons to produce UHE neutrinos. Further details on the GRB blast wave model and neutrino production model used in this paper can be found in \citet{Razzaque2013PhRvD..88j3003R, Razzaque2015PhRvD..91d3003R}.

\subsection{GRB population}
To understand how a large population of lGRBs contributes to the diffuse astrophysical UHE neutrino flux, their distributions throughout the Universe need to be considered. We adopted the redshift-luminosity distribution obtained from the \textit{Swift} and \textit{Fermi}-GBM observations of lGRBs \citep{Banerjee2021ApJ...921...79B} to construct the evolution function for the lGRB population. The distributions of lGRBs in gamma-ray luminosity and redshift are independent of one another, i.e. $\Psi(L_\gamma, z) = \psi(L_\gamma)\rho(z)$, where $\psi(L_\gamma)$ and $\rho(z)$ denote the luminosity function and redshift distribution, respectively. The luminosity function $\psi(L_\gamma)$ is modelled as a broken power law and is given by 
\begin{equation}
\label{eq:luminosity_function}
    \psi(L_\gamma)=c_0\left[ \left(\cfrac{L_\gamma}{L_0}\right)^{-\alpha} + \epsilon\left(\cfrac{L_\gamma}{L_0}\right)^{-\beta} \right],
\end{equation}
where $\alpha=1.33$, $\beta=1.42$, and $\epsilon=1$, and the break luminosity is $L_0=3\times 10^{53}$ erg s$^{-1}$. The normalisation constant $c_0$ is fixed by integrating this expression over the considered luminosity range: $c_0=[\int \psi(L_\gamma)dL]^{-1}$. The redshift distribution function is
\begin{equation}
    \label{eq:redshift_distribution}
    \rho(z) = \rho_0\cfrac{(1+z)^{2.7}}{1+\left(\frac{1+z}{2.9}\right)^{5.6}},
\end{equation}
with $\rho_0=5.5$ Gpc$^{-3}$ yr$^{-1}$. The values for both expressions are the best-fit values from \citet{Banerjee2021ApJ...921...79B}. The luminosity ($L_\gamma$) considered is the intrinsic isotropic luminosity as inferred from observed GRBs in the $8$~keV--$40$~MeV \textit{Fermi}-GBM  and \textit{Swift} energy range. The lower and upper bounds we used for the luminosity function are $10^{49}$~erg~s$^{-1}$ and $10^{54}$~erg~s$^{-1}$, and we integrated the lGRB population out to redshift $z_{\rm max}=5$, as the number of GRBs decreases significantly at high redshift, and the contributions from far-away GRBs (z>5) to a diffuse flux become negligible.

\subsection{Diffuse neutrino flux}
The final step in our model consists of combining the individual neutrino fluxes from long GRBs with their cosmological evolution and gamma-ray luminosity distribution. The total diffuse flux expected on Earth is
\begin{eqnarray}
    \label{eq:total_flux}
    \Phi_\nu^{\rm tot}(E_\nu, E_k, n_0; \theta) = \int_0^{z_{\rm max}} \frac{1}{1+z}\frac{dV}{dz} ~~~~~~~~~~~~~~~~~~~~~~~~~~~~~~~\\ \nonumber 
    \times \int_{L_1}^{L_2}\Psi(L_\gamma, z) S_{\!\nu}(E_\nu, E_k, n_0; \theta) dL_\gamma dz,
\end{eqnarray}
where the $1/(1+z)$ factor corrects for the time dilation between distant lGRBs and the measured flux. The term $S_{\!\nu}$ is the neutrino fluence, i.e. integrated flux of all flavours over a time $t \ge t_{\rm dec}$ for a GRB of luminosity ($L_\gamma$) at a redshift $z$, where an adiabatic blast wave is evolving in the ISM of constant density $n_0$. We denote all other model parameters with $\theta$.
A similar expression holds for a wind-type medium by replacing $n_0$ with the wind parameter $A_*$. A detailed calculation of $S_{\!\nu}$ can be found in \citet{Razzaque2013PhRvD..88j3003R} and in Appendix A. To compute the diffuse flux, we integrated over the comoving volume. The differential comoving volume element is given by
\begin{equation}
    \label{eq:como_vol}
    \frac{dV}{dz}=\frac{4\pi c}{1+z}\left| \frac{dt}{dz}\right|d_L^2,
\end{equation}
where $d_L$ is the luminosity distance~\citep{Hogg:1999ad}. The cosmic time, $dt/dz$, is given by
\begin{equation}
    \label{eq:cosmic_time}
    \frac{dt}{dz}=\frac{-1}{H_0(1+z)\sqrt{\Omega_m(1+z)^3+\Omega_\Lambda}}.
\end{equation}
For the calculations in this work, we used $\Omega_m=0.286$, $\Omega_\Lambda=1-\Omega_m=0.714$, and $H_0=69.32$ km~s$^{-1}$~Mpc$^{-1}$~\citep{2014A&A...571A..16P}.

\section{Analysis method}
To calculate the total diffuse neutrino flux from the GRB blast wave models outlined above, we used the nested sampling algorithm as implemented in the \texttt{UltraNest} package~\citep{ultranest}. In this approach, the parameter-free \textit{MLFriend} algorithm is applied to sample a constrained likelihood and iteratively converge on the global maximum through a bootstrapping method~\citep{2016S&C....26..383B, 2019PASP..131j8005B}. By defining the total integrand of our diffuse neutrino flux as the likelihood, the integration is performed by Bayesian inference with the aforementioned nested sampling. The computed marginal likelihood, also known as evidence, corresponds to the evaluated integral, i.e. the total diffuse neutrino flux. With this approach, we continuously varied all integration parameters associated with the diffuse flux calculation without making any a priori assumptions.

The most likely diffuse UHE neutrino flux responsible for producing KM3-230213A has been calculated and reported in~\citet{KM3NeT:2025global}. By considering that no events have been reported in the IceCube Extremely-High-Energy (IC-EHE) or sensitive Auger selections, the UHE neutrino flux of a single flavour was estimated to be $E^2 \Phi^{\rm 1f}_{\nu + \bar\nu} = (7.5^{+13.1}_{-4.7}) \times 10^{-10}~{\rm GeV~cm^{-2}~s^{-1}~sr^{-1}}$ in the 90\% reconstructed energy range~\citep{Nature}. To constrain the baryon loading and the density parameters of the GRB blast wave model, we repeated the calculation of~\citep{KM3NeT:2025global} by including the total exposures from IC-EHE and Auger. The total number of expected events of a single flavour from the diffuse GRB flux for the combined three experiments is
\begin{equation}
    \label{eq:nexp}
    n_{\rm exp}(E_k, n_0; \theta) = \frac{4\pi}{3}\int_{E_{\rm min}}^{E_{\rm max}} \mathcal{E}^{\rm tot}
    \Phi_\nu^{\rm tot}(E_\nu, E_k, n_0; \theta)dE_\nu,
\end{equation}
where $\mathcal{E}^{\rm tot}=\sum_dT^dA_{\rm eff}^d$ is the summed total exposure of all three experiments. The term $T$ denotes the total lifetime associated with the selection, and $A_{\rm eff}$ is the corresponding effective area. The running index is $d\in$\{KM3NeT, IC-EHE, Auger\}, and the factor $1/3$ corrects for considering all three neutrino flavours both in the diffuse flux and in the effective areas used. We set the integration limits as $E\in[1 ~{\rm GeV}, 100~{\rm EeV}]$. The KM3NeT exposure $\mathcal{E}^{\rm KM3NeT}$ is associated with the bright track selection reported in~\citet{Nature}. The corresponding lifetime is 335 days, associated with the 19 and 21 detection unit KM3NeT/ARCA detector configuration, out of the foreseen 230 days \citep{KM3NeT:2024paj}, and the effective area is all-flavour sky-averaged and averaged between neutrinos and anti-neutrinos. The IC-EHE exposure $\mathcal{E}^{\rm IC-EHE}$ is extracted from the 9-year analysis by \citet{IceCube:2018fhm} with a lifetime of 3145.5 days. The Pierre-Auger sky-averaged exposure, $\mathcal{E}^{\rm Auger}$, is computed in~\citet{KM3NeT:2025global} by considering the Earth-skimming, low-zenith downward-going, and high-zenith downward-going samples with the effective area from the data release \citep{PierreAuger:2019azx}. The corresponding lifetime is 6574.5 days, taken from the data release \citep{PierreAuger:2023pjg}. No neutrino events have been reported above tens of PeV in either the IC-EHE or Auger analyses. As we are using publicly available data, the exact treatment of systematics and the potential background of these analyses have been omitted.

We constructed the likelihood by considering the simple Poisson probability of observing one event in any of the three experiments given the predicted number of events from the considered GRB model:
\begin{equation}
    \label{eq:likelihood}
    \mathcal{L}(E_k, n_0; \theta)= {\rm Poisson}\,(n_{\rm obs}; n_{\rm exp} (E_k, n_0; \theta)).
\end{equation}
We performed a 2D scan by varying the two parameters of interest, $f_b\eta\in[10^{-2},10^{1.7}]$, which corresponds to varying the ratio of kinetic energy to gamma-ray luminosity through Eq.~(\ref{eq:ratio}) and $n_0\in[1, 100]$ while keeping all other model parameters fixed. We then calculated the likelihood as defined in Eq.~(\ref{eq:likelihood}) at each point in this parameter space.

\begin{figure}[hbtp]
    \centering
    \includegraphics[width=\hsize]{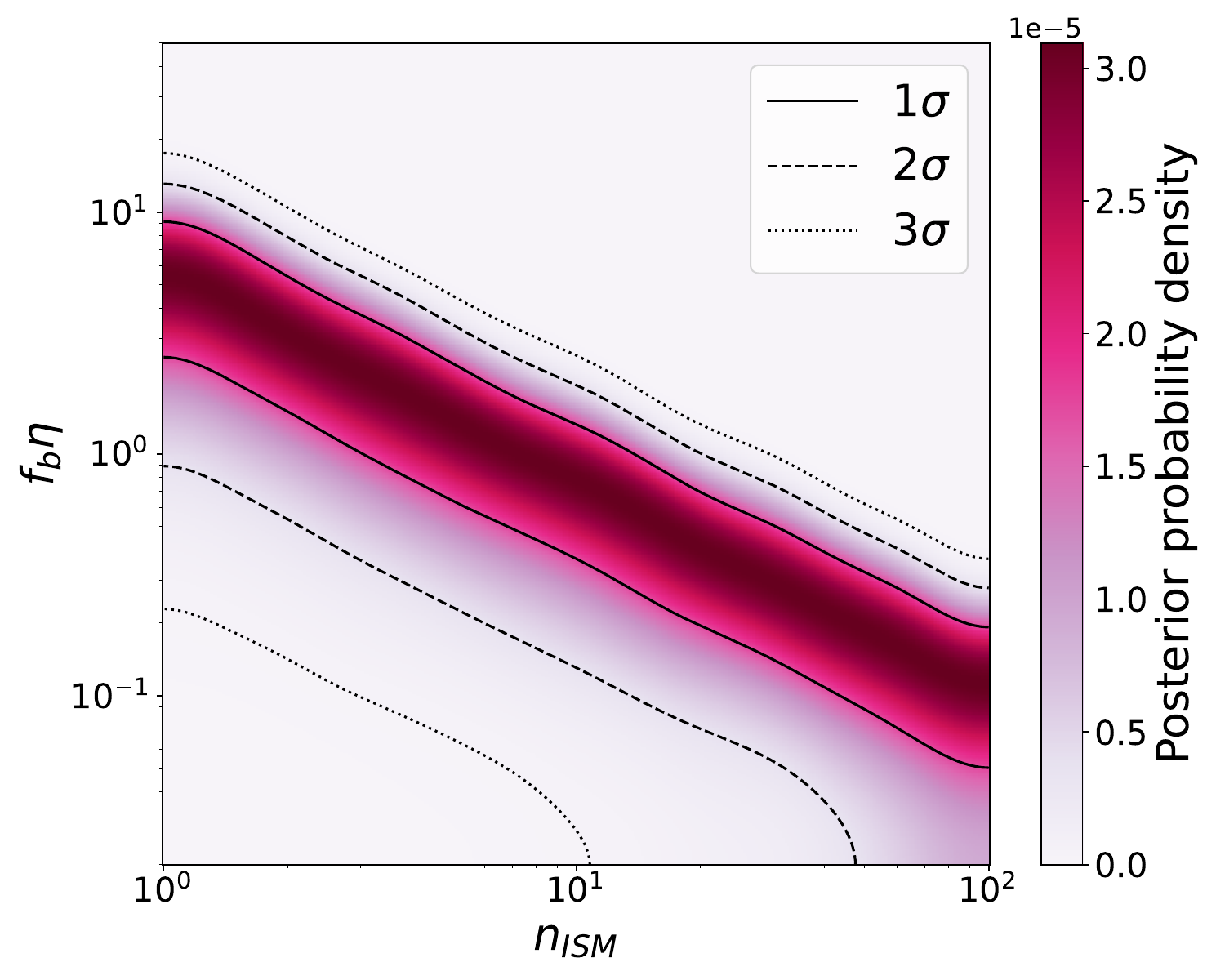}
    \caption{Posterior probability density of the 2D scan in $f_b\eta$ and $n_0$ for the ISM model. The solid (dashed, dotted) lines show the $1\sigma$ ($2\sigma$, $3\sigma$) contours.}
    \label{fig:2d_ppd}
\end{figure}

To constrain the two scan parameters, we adopted the Bayesian interpretation of probability and defined the posterior probability density as
\begin{equation}
    \label{eq:ppd}
    P(f_b\eta,n_0;\theta) = P(E_k/L_\gamma, n_0; \theta) =  a_n\mathcal{L}(E_k/L_\gamma, n_0; \theta),
\end{equation}
where we introduced the normalisation constant $a_n$ such that the full posterior parameter space integrates to unity. We assumed uniform priors on both varying parameters.

The corresponding 1D marginalised probability distributions can be used to extract the best-fit and confidence intervals. Our results do not indicate well-defined constraints on the parameters. This is caused by the high degeneracy in the posterior distribution, which can be seen from Fig.~\ref{fig:2d_ppd} and the intrinsic uncertainty in the convergence of the nested sampling algorithm. Since the expected number of events increases with both parameters of interest $(f_b\eta,n_0)$, we calculated the conditional posterior probability density as
\begin{equation}
    \label{eq:transformed_ppd}
    p(f_b\eta|n_0) = \frac{p(f_b\eta|n_0)}{\int p(f_b\eta,n_0)df_b\eta}.
\end{equation}
The resulting conditional probability density functions are shown in Fig.~\ref{fig:1d_ppd}.

\section{Results and discussion}
In order to calculate the diffuse flux from the population of GRB blast waves while varying both the baryon loading and the density of the surrounding medium, we fixed the remaining model parameters in $\theta$. This procedure is described in Appendix A, and the corresponding values are listed in Table~\ref{table:fixed_parameters}.

\begin{table}[hbtp]
\caption{Fixed GRB blast wave model parameters.}
\label{table:fixed_parameters}
\centering
\begin{tabular}{c | c c c c c c c c}
\hline\hline
GRB model & $\Gamma_0$ & $\epsilon_B$ & $\epsilon_e$ & $\epsilon_p$ & $\phi$ & $k$ & $a$ \\
\hline
   ISM & $10^{2.8}$ & $0.01$ & $0.1$ & $1.0$ & $10.0$ & $2.5$ & $4$ \\
   WIND & $10^{2.8}$ & $0.01$ & $0.1$ & $1.0$ & $10.0$ & $2.5$ & $4$ \\
\hline
\end{tabular}
\end{table}

\begin{figure}[hbtp]
    \centering
    \includegraphics[width=
    \hsize]{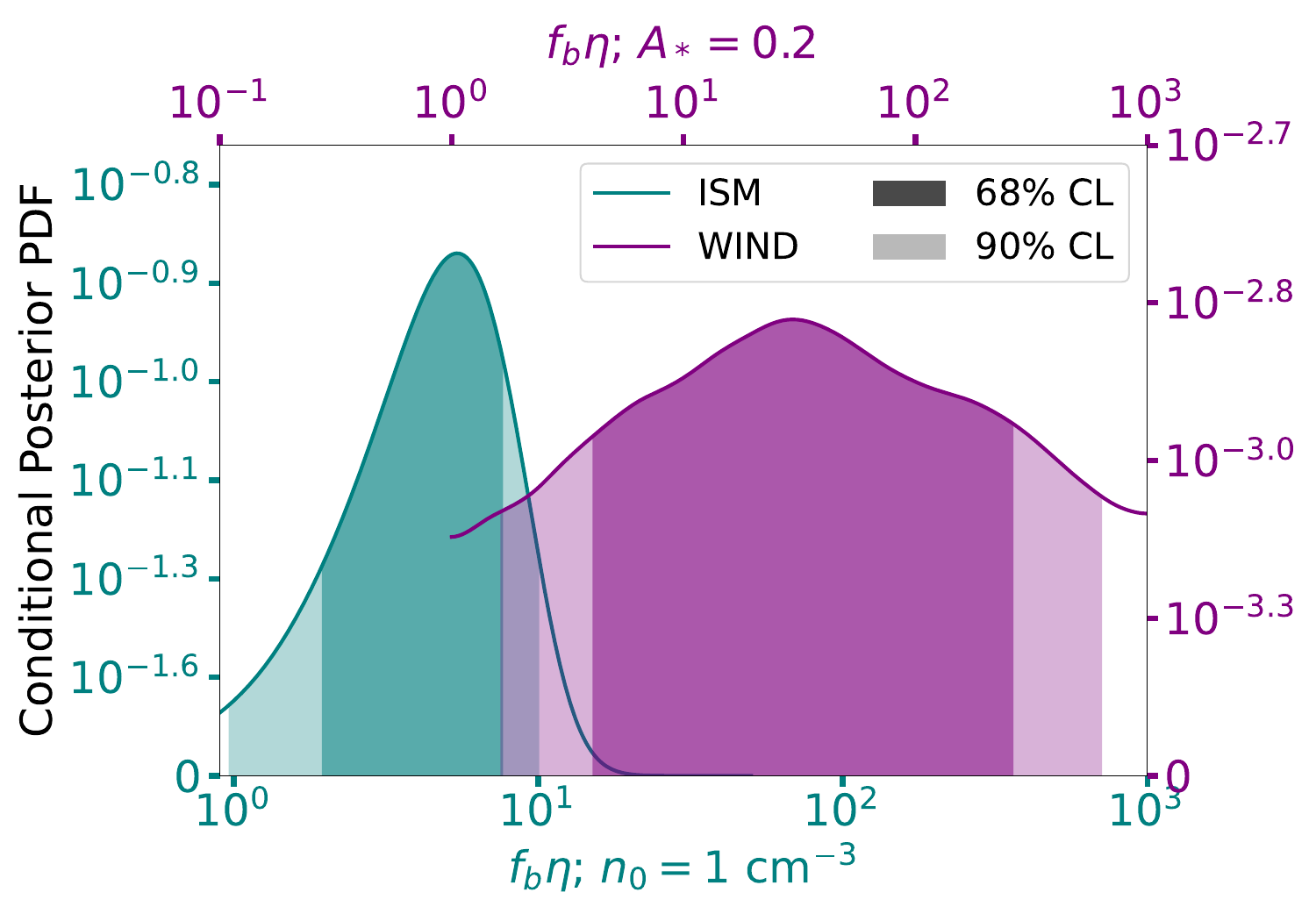}
    \caption{Conditional posterior probability density for $f_b\eta$ calculated by fixing the density parameter to $n_0=1$ cm $^{-3}$ ($A_*=0.2$) for the ISM (WIND) model. The solid lines show the distribution of the conditional posterior, and the dark (light) shaded area shows the $68\%$ ($90\%$) confidence region. The ISM blast wave model is indicated by the teal colour and corresponding lower and left axes. The WIND blast wave model is shown in purple with the right and upper axes. For the ISM model, the best fit and $90\%$ confidence level is $f_b\eta=5.4^{+4.7}_{-4.4}$. For the WIND model, it is $f_b\eta=29.5^{+609}_{-28}$.}
    \label{fig:1d_ppd}
\end{figure}

\begin{figure*}[hbtp]
\centering
\includegraphics[width=17cm]{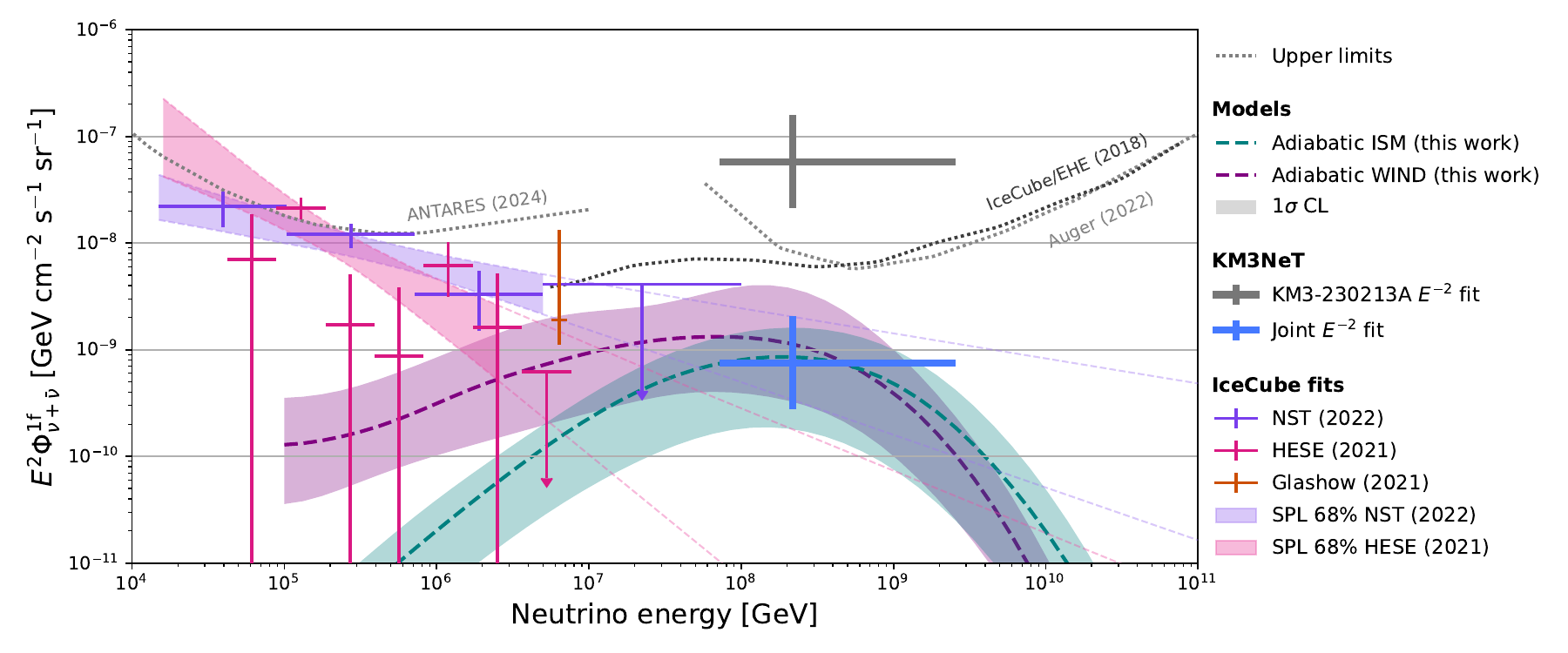}
      \caption{Energy-squared per-flavour diffuse astrophysical neutrino flux assuming ($\nu_e\!:\!\nu_\mu\!:\!\nu_\tau\!=\!1\!:\!1\!:\!1$) flavour equipartition. The teal (purple) dashed line corresponds to the diffuse flux from the ISM (WIND) blast wave models described in the text, with the corresponding 68\% confidence interval. The reported KM3-230213A flux~\citep{Nature} and corresponding 90\% neutrino energy range is indicated by the grey cross. The joint fit flux considering non-observation in the IceCube-EHE and Auger samples in the same energy range~\citep{KM3NeT:2025global} is shown by the blue cross. The 68\% confidence level contours from the IceCube NST~\citep{IceCube:2021uhz} and HESE~\citep{IceCube:2020wum} diffuse flux analyses are shown with the magenta and purple contours, respectively. The corresponding segmented fit analyses are shown by the magenta and purple crosses, and the IceCube Glashow resonance event~\citep{IceCube:2021rpz} is shown with an orange cross. The dotted lines show the upper limits from the ANTARES~\citep{ANTARES:2024ihw}, IceCube-EHE~\citep{IceCube:2018fhm}, and Auger~\citep{PierreAuger:2023pjg} analyses.}
      \label{fig:spectral_plot}
\end{figure*}

The $\Gamma_0$ parameter was fixed to a constant, as it determines the deceleration time ($t_{\rm dec}$) of the model, which serves as the lower-limit of the time-integration over flux in Eq.~(\ref{eq:fluence}). The parameter $\epsilon_e = 0.1$ and $\epsilon_B = 0.1-0.01$ are standard choices in GRB afterglow modelling \citep[see e.g.][]{Kumar2015, Miceli2022} based on particle-in-cell simulations \citep{Sironi2011, Sironi2013}. Additionally, the value of $\epsilon_B$ was chosen after performing a parameter space scan in $f_b\eta$ versus $\epsilon_B$ similar to that described above while fixing $n_0=1$ cm$^{-3}$. The resulting posterior is shown in Fig. \ref{fig:appb_ppd_fbeB}. This yielded a best-fit value of $\epsilon_B$ that is degenerate with $f_b\eta$, but it agrees well with broad-band modelling of very-high-energy GRBs \citep{Barnard:2025zjx}.
Using the conditional posterior from Eq.~(\ref{eq:transformed_ppd}), we estimated the best-fit value of $f_b\eta$ that corresponds to one observed event in the combined exposure of KM3NeT, IceCube-EHE, and Auger. The best fit with its associated 90\% confidence level is found to be $f_b\eta = 5.4^{+4.7}_{-4.4}$ for fixed $n_0=1$ cm$^{-3}$. These values were extracted from the conditional posteriors show in Fig. \ref{fig:1d_ppd}, where we observed a clear highest posterior density point defining the best-fit value and a smooth distribution covering the 68\% and 90\% credible interval. The diffuse neutrino flux, assuming these best-fit parameters, and the corresponding 68\% containment region, is shown in Fig.~\ref{fig:spectral_plot}. Fixing the efficiency parameter to $\eta=0.2$, we constrained the baryon loading to be $f_b\leq 51$ at a 90\% confidence level for the ISM model.

For the WIND model, we find the best-fit value and 90\% confidence interval to be $f_b\eta=29.5^{+609}_{-28}$ for $A_*=0.2$. This corresponds to a baryon loading of $f_b\approx148$. Conversely, when fixing $f_b=50$, we found $A_*=0.17^{+0.08}_{-0.10}$ at a 90\% confidence level. For this model, the density parameter, $A_*$, has a significantly larger impact on the diffuse flux than the baryon loading. We also note that increasing the two parameters simultaneously does not necessarily give a higher flux. The GRB blast wave in the WIND model loses significant amounts of energy through adiabatic expansion before the deceleration time, where UHE neutrino production proceeds photo-hadronically, and thus the flux is reduced. If the density of the surrounding material is too high or the baryon loading is significant, too much of the kinetic energy in the blast wave is lost in the initial interactions with the surrounding medium. As a consequence, the remaining shock energy will not be sufficient to accelerate the protons to UHE, reducing the subsequent UHE neutrino flux. This is not the case for the ISM mode, where the UHE neutrino fluence increases with $f_b$.

From the analytical expressions in Appendix A, we observed that for the ISM model, the time-dependent bulk Lorentz factor $\Gamma$ is independent of $\Gamma_0$. Thus, the initial bulk Lorentz factor only enters the expression for the deceleration time; varying the value of $\Gamma_0$ has a negligible impact on the total UHE neutrino flux (see Fig. \ref{fig:appb_ppd}). For the WIND model, the $\Gamma_0$ parameter does influence the UHE neutrino flux.

In both of the GRB blast wave models, we only considered neutrinos produced photo-hadronically by the interaction of accelerated UHE protons in the jet with photons from the afterglow. These $p\gamma$ interactions dominate over the hadro-nuclear ($pp, pn)$ interactions due to the significantly diluted particle density in the environment surrounding the GRB progenitor~\citep{Waxman1997PhRvL..78.2292W, IC_GRB_2016ApJ...824..115A}. Moreover, we only considered the primary single-pion production channel in the $p\gamma$ interactions. The authors of references~\citet{Murase2007PhRvD..76l3001M} and \citet{Razzaque2013PhRvD..88j3003R} have compared the full pion-production cross-section while considering higher multiplicities, and they found that the contribution is small below $E_p\sim10^{18}$ eV.

In this analysis, we chose to fit the product of baryon loading and the efficiency parameter $\eta$ rather than the baryon loading directly. They are related by Eq.~(\ref{eq:ratio}). Thus, our results are interpretable for other values of the efficiency parameter $\eta$.

Previous analyses constrained the baryon loading from non-observations of sub-PeV neutrinos~\citep[see e.g.][]{IC_GRB_2016ApJ...824..115A}. These analyses primarily focused on the neutrino emission from the GRB prompt phase and not from the subsequent afterglow. In the models investigated by~\citet{IC_GRB_2016ApJ...824..115A}, the baryon loading factor $f_b$ varies as a function of the Lorentz factor $\Gamma_0$. In the internal shock model considered therein, $f_b$ is found to be $\lesssim 10$ at a 90\% confidence level, which is more constraining than our result. However, this is only for $\Gamma_0\lesssim 300$. Since the considered ISM model does not depend strongly on $\Gamma_0$ (see Appendix B), when comparing with the value $\Gamma_0=10^{2.8}$ used for our calculations, the 90\% confidence level from~\citet{IC_GRB_2016ApJ...824..115A} is $f_b\lesssim 200$. For the internal collision-induced magnetic reconnection and turbulence model considered therein~\citep{Zhang_2010}, the constraints are consistent with our findings also for lower $\Gamma_0$ values. The photospheric model, which is severely constrained in~\citet{IC_GRB_2016ApJ...824..115A}, is disfavoured by the non-observation of a dominant thermal component in the GRB spectra~\citep{GBM_catalog_2012ApJS..199...19G}. Finally, we note that the baryon loading factor may be different in the initial prompt emission phase of the GRBs and the afterglow. Although the fraction of baryons in the prompt phase will contribute to the particle spectra in the afterglow emission, interactions with the surrounding environment and hadro-nuclear interactions in the prompt phase itself can alter the baryon-to-electron energy ratio during the evolution of the jet. Furthermore, if the GRB jet is dominated by the magnetic field \citep{Lyutikov2003astro.ph.12347L}, there will be fewer baryons to produce neutrinos in the prompt phase. Constraints on $f_b$ from the prompt (afterglow) phase may therefore not directly relate to the afterglow (prompt) emission.

\citet{IC_GRB_2022ApJ...939..116A} found that the prompt emission of GRBs observed in the electromagnetic spectrum contributes to less than 1\% of the observed diffuse neutrino flux below 10 PeV and that emission from the afterglow on timescales up to $10^4$ s is $\lesssim24\%$. Although our analysis focuses on the $\gtrsim 10$ PeV energy range associated with KM3-230213A, we see from Fig.~\ref{fig:spectral_plot} that the diffuse neutrino flux below 10 PeV for both models is well below that of the NST and HESE IceCube fits. Our results are consistent with \citet{IC_GRB_2022ApJ...939..116A} regarding the contribution of GRBs to the diffuse neutrino flux at lower energies and allow for the interactions of lGRB blast waves to explain the UHE diffuse neutrino flux.

Although our focus has been exclusively on GRBs, other classes of astrophysical objects, such as active galactic nuclei (AGNs) may also be significant contributors to the diffuse high-energy neutrino flux. Luminous AGNs, such as blazars, are expected to accelerate protons to UHEs in highly collimated jets, resulting in high-energy neutrino production when these protons interact with the varying radiation fields. As the predominant neutrino production channel is through pion decay, significant emission of high-energy gamma-rays from $\pi^0$ decays will accompany neutrinos from such non-transient sources as blazars. The measurement of the diffuse gamma-ray sky by \textit{Fermi}-LAT~\citep{atwood2009large} places strong constraints on the contribution of steady sources to the diffuse UHE neutrino flux. Since GRBs are transient sources, they are not affected by these constraints. 
Other types of transient sources are also candidates for UHE neutrino production, such as tidal disruption events~\citep[e.g.][]{lunardini2017high} and magnetar-powered super-luminous supernovae~\citep[e.g.][]{Fang:2018hjp}. However, the current sparsity of their detection in gamma rays makes the population and cosmic evolution of these transients less known and their contribution to the diffuse UHE neutrino flux uncertain~\citep[e.g.][]{Das2025arXiv250410847D}.

\section{Conclusions}
We have calculated the contribution of different GRB blast wave models to the diffuse neutrino flux and constrained the lGRB model parameters with respect to the diffuse UHE neutrino flux most likely associated with KM3-230213A~\citep{Nature}. The total neutrino fluence from individual lGRBs was calculated following the blast wave models presented by~\citet{Razzaque2013PhRvD..88j3003R} and the lGRB luminosity function used was fitted to \textit{Swift} and \textit{Fermi}-GBM observations by~\citet{Banerjee2021ApJ...921...79B}. The total diffuse flux was calculated using the nested sampling algorithm implemented in the \texttt{UltraNest} python package, allowing the variation of all integration model parameters, which takes into consideration a broader population of lGRBs up to $z=5$ following the fitted luminosity function.

We considered two different models for UHE neutrino production from GRB blast waves: one in which the density of the surrounding matter remains constant around the GRB progenitor and another in which the density decreases radially. For the GRB blast wave model with constant ISM density $n_0=1$ cm$^{-3}$, the baryon loading is constrained to be $f_b\leq 51$ at 90\% confidence. Assuming a larger value for the ISM density, $f_b$ is significantly more constrained. The corresponding best-fit baryon loading, $f_b=27$, would produce a diffuse UHE neutrino flux consistent with the detection of one UHE event (i.e. KM3-230213A) within the cumulative exposure of KM3NeT, IceCube-EHE and Auger. When assuming the GRB blast wave interacts with a wind-type medium with radially decreasing density, the baryon loading is less well defined, as the density parameter ($A_*$) dominates. It is constrained to be $f_b\leq 1065$ at 90\% confidence for $A_*=0.1$. By fixing the baryon loading factor to $f_b\sim50$, i.e. typical values from prompt emission models \citep{IC_GRB_2016ApJ...824..115A}, we found the best-fit value of the density parameter $A_*=0.017^{+0.08}_{-0.10}$ at 90\% confidence.

Our results show that a large population of lGRBs can give rise to the diffuse UHE neutrino flux associated with KM3-230213A. Moreover, both GRB models we considered are shown to be consistent with existing limits on their contribution to the diffuse neutrino flux at lower energies. Although the true diffuse neutrino flux at UHEs may come from additional sources~\citep[see][]{meszaros2017astrophysical}, GRB blast waves can contribute significantly to the UHE neutrino flux required for KM3-230213A whilst remaining consistent with previous limits on GRB model parameters \citep{IC_GRB_2016ApJ...824..115A}. Future observations by upcoming large radio detectors such as GRAND \citep{alvarez2020giant}, Askaryan detectors such as RNO-G \citep{aguilar2021design}, and combined Askaryan and Cherenkov detectors such as IceCube-Gen2 \citep{IceCube-Gen2:2020qha} can contribute to better characterisation of the UHE neutrino flux. Our modelling of the UHE neutrino flux corresponding to KM3-230213A motivates further observations of the electromagnetic sky and exploration of the role of GRBs as multi-messenger sources.

\begin{acknowledgements}
We thank the anonymous reviewers for their insightful comments and constructive suggestions, which greatly improved this manuscript.
The authors acknowledge the financial support of:
KM3NeT-INFRADEV2 project, funded by the European Union Horizon Europe Research and Innovation Programme under grant agreement No 101079679;
Funds for Scientific Research (FRS-FNRS), Francqui foundation, BAEF foundation.
Czech Science Foundation (GAČR 24-12702S);
Agence Nationale de la Recherche (contract ANR-15-CE31-0020), Centre National de la Recherche Scientifique (CNRS), Commission Europ\'eenne (FEDER fund and Marie Curie Program), LabEx UnivEarthS (ANR-10-LABX-0023 and ANR-18-IDEX-0001), Paris \^Ile-de-France Region, Normandy Region (Alpha, Blue-waves and Neptune), France,
The Provence-Alpes-Côte d'Azur Delegation for Research and Innovation (DRARI), the Provence-Alpes-Côte d'Azur region, the Bouches-du-Rhône Departmental Council, the Metropolis of Aix-Marseille Provence and the City of Marseille through the CPER 2021-2027 NEUMED project,
The CNRS Institut National de Physique Nucléaire et de Physique des Particules (IN2P3);
Shota Rustaveli National Science Foundation of Georgia (SRNSFG, FR-22-13708), Georgia;
This work is part of the MuSES project which has received funding from the European Research Council (ERC) under the European Union’s Horizon 2020 Research and Innovation Programme (grant agreement No 101142396).
The General Secretariat of Research and Innovation (GSRI), Greece;
Istituto Nazionale di Fisica Nucleare (INFN) and Ministero dell’Universit{\`a} e della Ricerca (MUR), through PRIN 2022 program (Grant PANTHEON 2022E2J4RK, Next Generation EU) and PON R\&I program (Avviso n. 424 del 28 febbraio 2018, Progetto NRRP), Italy; IDMAR project Po-Fesr Sicilian Region az. 1.5.1; A. De Benedittis, W. Idrissi Ibnsalih, M. Bendahman, A. Nayerhoda, G. Papalashvili, I. C. Rea, A. Simonelli have been supported by the Italian Ministero dell'Universit{\`a} e della Ricerca (MUR), Progetto CIR01 00021 (Avviso n. 2595 del 24 dicembre 2019); KM3NeT4RR MUR Project National Recovery and Resilience Plan (NRRP), Mission 4 Component 2 Investment 3.1, Funded by the European Union – NextGenerationEU,CUP I57G21000040001, Concession Decree MUR No. n. Prot. 123 del 21/06/2022;
Ministry of Higher Education, Scientific Research and Innovation, Morocco, and the Arab Fund for Economic and Social Development, Kuwait;
Nederlandse organisatie voor Wetenschappelijk Onderzoek (NWO), the Netherlands;
The grant “AstroCeNT: Particle Astrophysics Science and Technology Centre”, carried out within the International Research Agendas programme of the Foundation for Polish Science financed by the European Union under the European Regional Development Fund; The program: “Excellence initiative-research university” for the AGH University in Krakow; The ARTIQ project: UMO-2021/01/2/ST6/00004 and ARTIQ/0004/2021;
Ministry of Education and Scientific Research, Romania;
Slovak Research and Development Agency under Contract No. APVV-22-0413; Ministry of Education, Research, Development and Youth of the Slovak Republic;
MCIN for PID2021-124591NB-C41, -C42, -C43 and PDC2023-145913-I00 funded by MCIN/AEI/10.13039/501100011033 and by “ERDF A way of making Europe”, for ASFAE/2022/014 and ASFAE/2022 /023 with funding from the EU NextGenerationEU (PRTR-C17.I01) and Generalitat Valenciana, for Grant AST22\_6.2 with funding from Consejer\'{\i}a de Universidad, Investigaci\'on e Innovaci\'on and Gobierno de Espa\~na and European Union - NextGenerationEU, for CSIC-INFRA23013 and for CNS2023-144099, Generalitat Valenciana for CIDEGENT/2020/049, CIDEGENT/2021/23, CIDEIG/2023/20, ESGENT2024/24, CIPROM/2023/51, GRISOLIAP/2021/192 and INNVA1/2024/110 (IVACE+i), Spain;
Khalifa University internal grants (ESIG-2023-008, RIG-2023-070 and RIG-2024-047), United Arab Emirates;
The European Union's Horizon 2020 Research and Innovation Programme (ChETEC-INFRA - Project no. 101008324).
\end{acknowledgements}

\bibliographystyle{aa} 
\bibliography{references}

\begin{appendix}
\section{Individual GRB neutrino fluence}
\label{appendix:A}
We provide a more detailed analytical summary of the UHE neutrino flux produced in a single GRB blast wave and the total energy released in UHE neutrinos. The deceleration timescale is
\begin{equation}
    \label{eq:t_dec}
    t_{\rm dec}(z, E_k)=\left[\frac{3E_k(1+z)^3}{64\pi n_0m_pc^5\Gamma_0^8}\right]^{1/3},
\end{equation}
where $E_k$ is the kinetic energy in the blast wave, $n_0$ is the number density of the ISM, $m_p$ is the proton mass, $c$ is the speed of light, and $\Gamma_0$ is the initial bulk Lorentz factor of the outflow. The bulk Lorentz factor after the deceleration timescale evolves in the constant density ISM as
\begin{equation}
    \label{eq:bulk_lorentz}
    \Gamma(t)=\Gamma_0\left(\frac{t_{\rm dec}}{4t}\right)^{3/8},
\end{equation}
and the radius and magnetic field strength of the blast wave increases correspondingly as
\begin{equation}
    \label{eq:radius}
    R(t)=\frac{2\Gamma^2(t)act}{1+z},
\end{equation}
and
\begin{equation}
    \label{eq:b_field}
    B'(t)=[32\pi\epsilon_B n_0m_pc^2]^{1/2}\Gamma(t),
\end{equation}
respectively. The constant $a=4$; $\epsilon_B$ denotes the fraction of the forward-shock energy from the blast wave converted into magnetic energy; and the timescale is $t>t_{\rm dec}$. In the WIND model, we instead consider the deceleration timescale
\begin{eqnarray}
    \label{eq:t_dec_wind}
    t_{\rm dec}(z, E_k) = \frac{E_k(1+z)}{16\pi Am_pc^3\Gamma_0^4},
\end{eqnarray}
and the bulk Lorentz factor as
\begin{equation}
    \label{eq:bulk_lorentz_wind}
    \Gamma(t)=\Gamma_0\left( \frac{t_{\rm dec}}{4t} \right)^{1/4},
\end{equation}
with $A=3.02\times 10^{35}A_*$ cm$^{-1}$. Consequently, the density of the surrounding medium in the WIND model is $n(R)=AR^{-2}$, whereas it is constant for the ISM model.

Within this blast wave model, electrons accelerated in the shock exhibit distinct behaviours in different regimes characterised by different electron Lorentz factors
\begin{align}
    \label{eq:regimes}
    & \gamma'_m=\epsilon_e\frac{m_p}{m_e}\Gamma(t), \\
    & \gamma'_c=\frac{6\pi m_e c(1+z)}{\sigma_TtB'^2(t)\Gamma(t)}, \\
    & \gamma'_s=\left[\frac{6\pi e}{\sigma_TB'(t)\phi}\right]^{1/2},
\end{align}
corresponding to minimum, cooling, and saturation electron states, respectively. The primed notation indicates the blast wave comoving frame. The $\epsilon_e$ parameter describes the kinetic energy in the blast wave converted into random electron energy, $m_e$ is the electron mass, $\sigma_T$ is the Thomson cross section, and $\phi$ is the number of gyroradii for the accelerated electrons. These electrons produce a flux of synchrotron photons as
\begin{equation}
    \label{eq:synch_flux}
    F_\nu=\frac{N_e}{4\pi d_L^2}\frac{P(\gamma')}{h\nu}\frac{\Gamma^2(t)}{(1+z)^2},
\end{equation}
where $N_e=(4/3)\pi R^3(t)n(t)$ is the total number of electrons in the blast wave and
\begin{equation}
    \label{eq:lum_distance}
    d_L(z)=\frac{c(1+z)}{H_0}\int_0^z\frac{dz^*}{\sqrt{\Omega_m(1+z^*)^3+\Omega_\Lambda}},
\end{equation}
is the luminosity distance. The synchrotron power of electrons with Lorentz factor $\gamma'$ is given by
\begin{equation}
    \label{eq:synch_pow_elec}
    P(\gamma')=\frac{c\sigma_T}{6\pi}B'^2\gamma'^2,
\end{equation}
and the characteristic synchrotron frequency is
\begin{equation}
    \label{eq:char_synch_freq}
    h\nu=\frac{3}{2}\frac{B'(t)}{B_Q}\gamma'^2(t)m_ec^2\frac{\Gamma(t)}{1+z},
\end{equation}
with normalising magnetic field strength $B_Q=4.41\times 10^{13}$ G.

\subsection*{$p\gamma$ interactions}
The density of observed synchrotron photons is parametrised in the different regimes and calculated as
\begin{equation}
    n'_\gamma(\epsilon')=\frac{2d_L^2(1+z)F_{\nu,m}}{r^2c\Gamma\epsilon'}\times\begin{cases}
    (\epsilon'_c/\epsilon'_m)^{-3/2}(\epsilon'/\epsilon'_c)^{-2/3} \ \ \  {\rm for} \ \ \ \epsilon'_a<\epsilon'<\epsilon'_c \\
    (\epsilon'/\epsilon'_m)^{-3/2} \ \ \ {\rm for} \ \ \ \epsilon'_c<\epsilon'<\epsilon'_m \\
    (\epsilon'/\epsilon'_m)^{-k/2-1} \ \ \ {\rm for} \ \ \ \epsilon_m'<\epsilon'<\epsilon'_s,
    \end{cases}
\end{equation}
where $k$ is the spectral index of the electron distribution, $m,c,s$ correspond to the different regimes, and $\epsilon'=(h\nu)'=h\nu(1+z)/\Gamma$. The rate of $p\gamma$ interactions for protons with Lorentz factor $\gamma'_p$ in the blast wave with this synchrotron photon field is
\begin{equation}
    \label{eq:scattering_rate}
    K_{p\gamma}(\gamma'_p)=\frac{c}{2\gamma'^2_p}\int_{\epsilon'_{\rm th}}^\infty d\epsilon'_r\epsilon'_r\sigma_{p\gamma}(\epsilon'_r)\int_{\epsilon'_r/(2\gamma'_p)}^\infty d\epsilon'\frac{n'_\gamma(\epsilon')}{\epsilon'^2},
\end{equation}
where $\epsilon'_r$ is the photon energy in the rest frame of the proton and $\epsilon'_{\rm th}=m_\pi c^2+m_\pi^2c^2/2m_p$ is the threshold for pion production. The cross-section for $p\gamma\to n\pi^+$ interactions is
\begin{equation}
    \label{eq:pg_xsection}
    \sigma_{p\gamma}(\epsilon'_r)=\sigma_0\Gamma_\Delta^2s^2\epsilon'^{-2}_r[\Gamma_\Delta^2s+(s-m_\Delta^2)^2]^{-1},
\end{equation}
with $s=m_p^2c^4+2\epsilon'_rm_pc^2$, $\sigma_0=3.11\times 10^{-29}$ cm$^2$, and $\Gamma_\Delta=0.11$ GeV is the width of the Delta resonance. Finally, we calculate the optical depth for $p\gamma$ interactions as
\begin{equation}
    \label{eq:optical_depth}
    \tau_{p\gamma}(\gamma'_p)=K_{p\gamma}(\gamma'_p)\frac{R(t)}{2ac\Gamma}=K_{p\gamma}(\gamma'_p)\frac{t\Gamma}{1+z},
\end{equation}
where $a$ is the same constant as in Eq.~\ref{eq:radius}.

\subsection*{Proton population}
The majority of the multi-messenger emissivity from an individual GRB blast wave is driven by the interactions of protons accelerated by the shocks in the blast wave. In this model, the differential number density of protons as a function of proton energy is given by
\begin{equation}
    \label{eq:n_protons}
    n(E_p)=\frac{\mathcal{E}_{\rm CR}}{VE_p^2\ln(\gamma'_{p,s}/\Gamma)},
\end{equation}
where $\gamma'_{p,s}$ is the saturation proton Lorentz factor given by
\begin{equation}
    \label{eq:proton_sat_lor_fact}
    \gamma'_{p,s}=\frac{eB'(t)}{\phi m_pc}\frac{t\Gamma}{1+z}.
\end{equation}
Here, $\phi$ is the number of gyroradii for the protons. The energy of the cosmic rays in the blast wave after the deceleration timescale $t_{\rm dec}$ is
\begin{equation}
\label{eq:cr_energy}
    \mathcal{E}_{\rm CR}=\frac{4}{3}\pi\epsilon_pn_0r^3(t)m_pc^2[\Gamma^2-1],
\end{equation}
with $\epsilon_p$ being the fraction of the blast wave energy that goes into proton acceleration. If this population of blast wave protons could escape their source environment as cosmic rays, their energy would be
\begin{equation}
    \label{eq:proton_cr_flux}
    E_{p,s}=\frac{m_pc^2\gamma'_{p,s}\Gamma}{1+z}.
\end{equation}
To avoid having a population of runaway high-energy protons, we introduce an exponential cut-off term $\propto\exp(-E_p/E_{p,s})$ to suppress the proton spectrum after saturation. Including this cut-off, if the protons could freely escape the blast wave environment and make it to Earth, their flux would be
\begin{equation}
    \label{eq:proton_flux}
    J_p(E_p)=\frac{c}{4\pi}\left(\frac{R}{d_L}\right)^2n(E_p)e^{-E_p/E_{p,s}}.
\end{equation}

\subsection*{Neutrino flux}
The protons accelerated within the GRB blastwave shocks will interact with the synchrotron photons and produce neutrinos through pion production and consequent decays
\begin{align}
    \label{eq:neutrino_channels}
    & \pi^+\to\mu^++\nu_\mu, \\
    & \mu^+\to e^++\nu_e+\bar\nu_\mu \nonumber,
\end{align}
resulting in the $1:2:0$ flavour ratio associated with photopion interactions. Although neutrinos are also produced through neutron beta decay, this process is subdominant and consequently ignored in our calculations. We assume an equal probability of producing $\pi^+$ and $\pi^0$ with mean inelasticity $\langle x\rangle\approx 0.2$, resulting in a pion flux
\begin{equation}
    \label{eq:pion_flux}
    J_\pi(E_\pi)\approx\frac{1}{\langle x\rangle}J_p\left(\frac{E_\pi}{\langle x\rangle}\right)\tau_{p\gamma}\left(\frac{E_\pi(1+z)}{m_pc^2\langle x\rangle\Gamma}\right).
\end{equation}
The consequent flux of muons is found by integrating over the kinematics of the process governed by the mass ratio as
\begin{equation}
    \label{eq:muon_flux}
    J_\mu(E_\mu)=\int_0^1\frac{dx}{x}f_{\pi^+\to\mu^+}(x)J_\pi\left(\frac{E_\mu}{x}\right),
\end{equation}
where $x=E_\nu/E_\pi$. The resulting neutrino fluxes from all relevant decay channels are calculated by considering similar mass ratios and integrating over the kinematics of the interactions
\begin{align}
    \label{eq:neutrino_flux}
    & J_{\nu_\mu}(E_\nu)=\int_0^1\frac{dx}{x}f_{\pi^+\to\nu_\mu}(x)J_\pi\left(\frac{E_\nu}{x}\right); \ \ \ x=\frac{E_\nu}{E_\pi}; \\
    & J_{\nu_e}(E_\nu)=\int_0^1\frac{dy}{y}\int_0^1\frac{dx}{x}f_{\mu^+\to\nu_e}(x,y)f_{\pi\to\mu}(x)J_\pi\left(\frac{E_\nu}{xy}\right); \nonumber \\
    & J_{\bar\nu_\mu}(E_\nu)=\int_0^1\frac{dy}{y}\int_0^1\frac{dx}{x}f_{\mu^+\to\bar\nu_\mu}(x,y)f_{\pi\to\mu}(x)J_\pi\left(\frac{E_\nu}{xy}\right), \nonumber
\end{align}
where $x=E_\mu/E_\pi$ and $y=E_\nu/E_\mu$ in the last two expressions. The scaling functions $f$ are parametrisations of the decays as derived in~\citet{lipari1993lepton}. The final step is to integrate the total neutrino flux from a single GRB blastwave over its duration. The emission happens only after the deceleration time $t_{\rm dec}$ and decreases as time increases,
\begin{equation}
    \label{eq:fluence}
    S_\nu(E_\nu)=\int_{t_{\rm dec}}^\infty J_\nu(E_\nu)dt.
\end{equation}
This is the fluence of an individual GRB blastwave and gives the total energy released in neutrinos from photopion interactions.

\section{Dependence on other model parameters}
\label{appendix:B}

In the main text, we focused on what happens to the predicted neutrino flux as we vary the baryon loading and the density of the surrounding medium, while fixing the remaining model parameters. It is also worth investigating how changing the other parameters influences the UHE neutrino flux. We see from Eq.~\ref{eq:ratio} that modifying the efficiency of energy conversion from kinetic energy to gamma-ray luminosity $\eta$ is directly proportional to changing the ratio between kinetic energy and gamma-ray luminosity $E_k/L_\gamma$ itself. Thus, we chose to constrain the product of the baryon loading and this efficiency parameter in the main text.

\begin{figure}[h]
    \centering
    \includegraphics[width=\hsize]{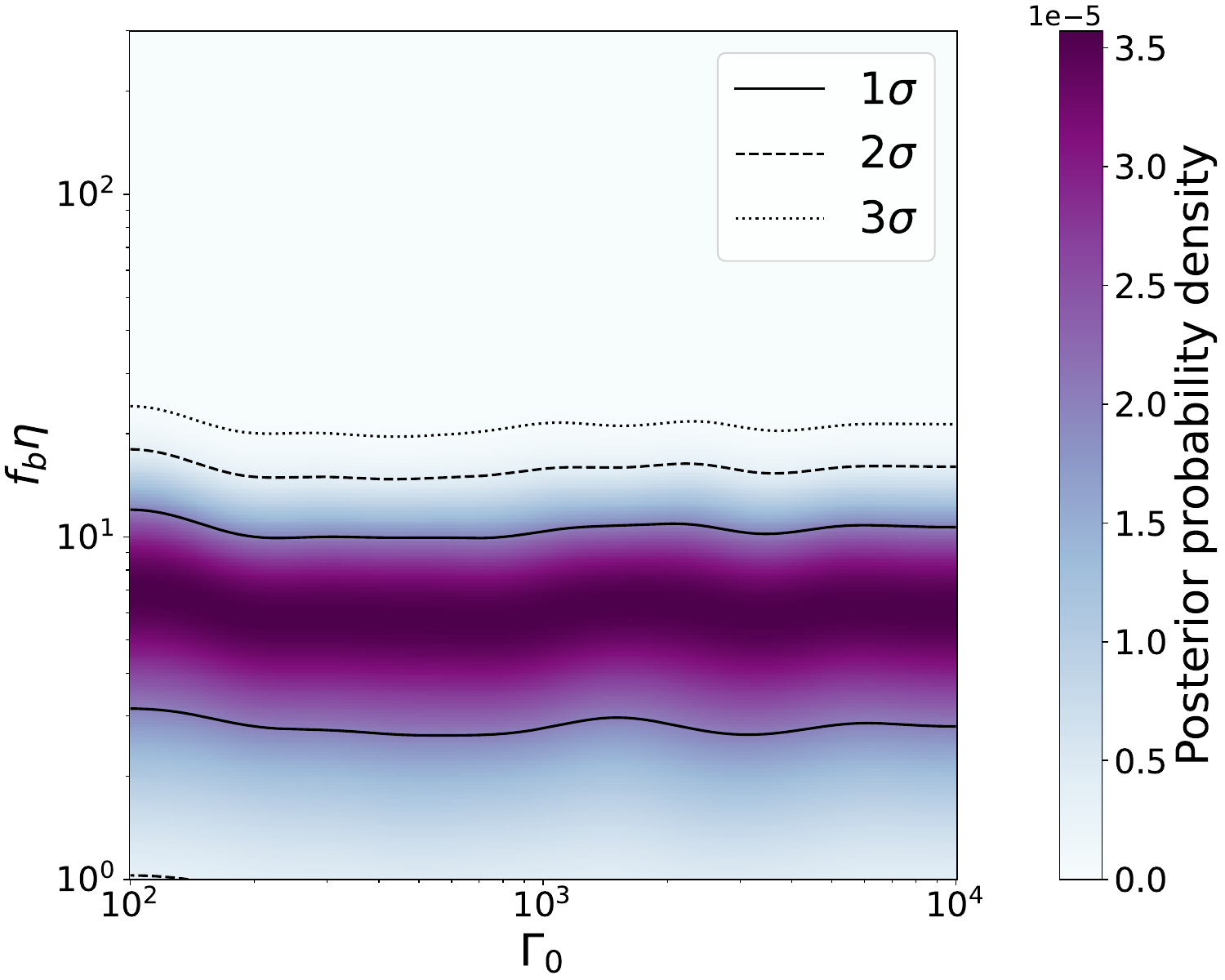}
    \caption{Posterior probability density of the 2D scan in $f_b\eta$ and $\Gamma_0$ for the ISM model with $n_0=1$ cm$^{-3}$. The solid (dashed, dotted) lines show the $1\sigma$ ($2\sigma$, $3\sigma$) contours.}
    \label{fig:appb_ppd}
\end{figure}

Since previous searches and literature focus on fitting the bulk Lorentz factor $\Gamma$, we also consider how this parameter influences our model and flux prediction. From the analytical expressions in Appendix A, we see that the model depends on the initial bulk Lorentz factor $\Gamma_0$, which determines the deceleration timescale $t_{\rm dec}$ and consequently the bulk Lorentz factor $\Gamma$. For the ISM model, the time-dependent bulk Lorentz factor $\Gamma(t)$ is independent of the initial bulk Lorentz factor $\Gamma_0$, and, since the model GRB flux only depends on $\Gamma(t)$ and not on $t_{\rm dec}$, the effects of varying $\Gamma_0$ are minimal. The deceleration timescale $t_{\rm dec}$ only enters the calculation via the lower integration bound of the single-GRB fluence of Eq.~\ref{eq:fluence}, and the impact on the diffuse UHE neutrino flux is small. This can be seen in Fig.~\ref{fig:appb_ppd}, where we fixed the value of the ISM density to $n_0=1$ cm$^{-3}$, and fitted instead for varying $\Gamma_0$. The same relation holds for the WIND model. In this case, the deceleration time $t_{\rm dec}$ depends more strongly on the initial bulk Lorentz factor, $t_{\rm dec}\propto\Gamma_0^{-4}$, and the time elapsed before UHE neutrino production is shorter. However, as this dependence only enters in Eq.~\ref{eq:fluence}, it does not significantly affect the diffuse UHE neutrino flux, especially considering the competing effect of reduction in the UHE neutrino flux with increasing $f_b$ as described in the main text.

\begin{figure}[h]
    \centering
    \includegraphics[width=\hsize]{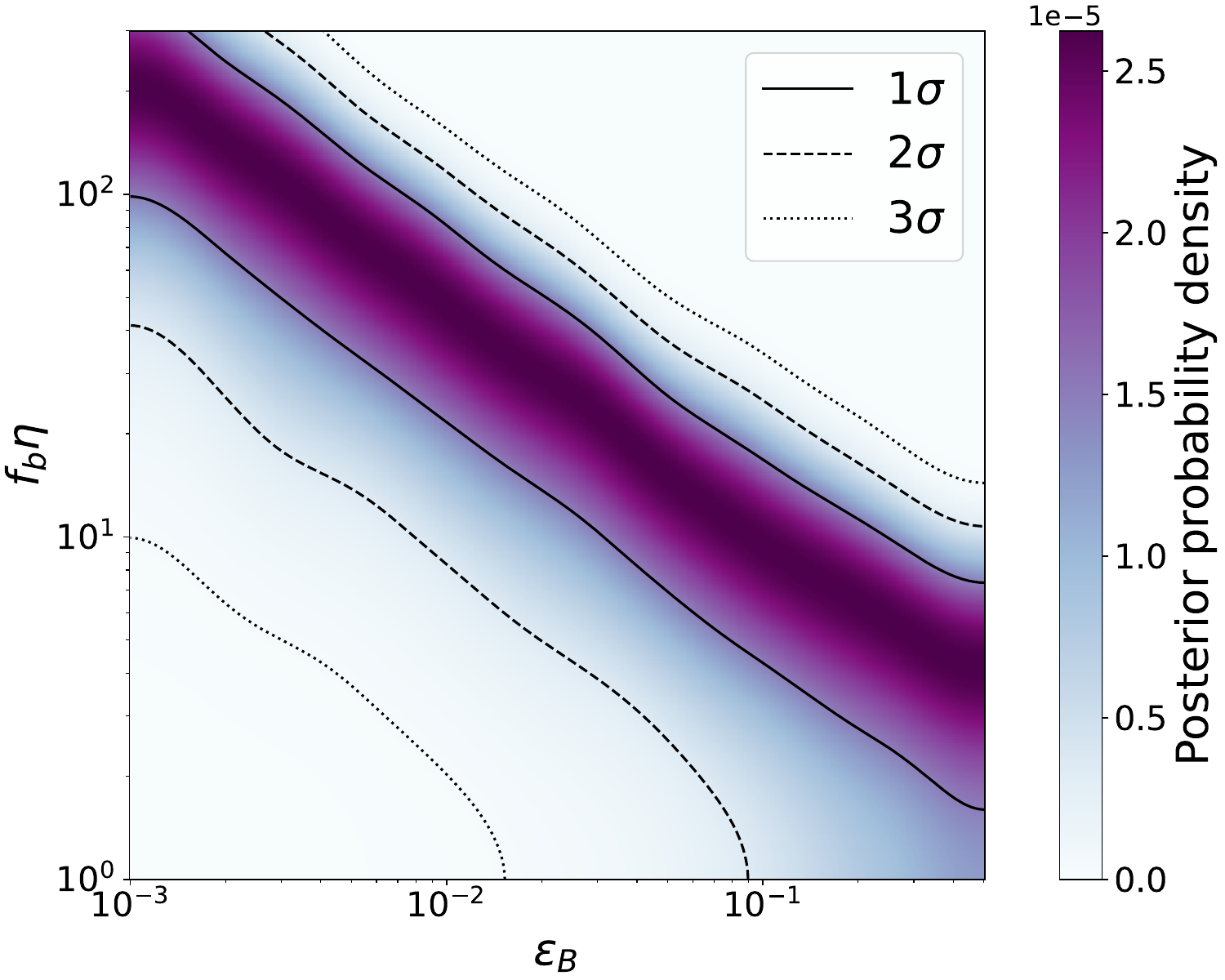}
    \caption{Posterior probability density of the 2D scan in $f_b\eta$ and $\epsilon_B$ for the ISM model with $n_0=1$ cm$^{-3}$. The solid (dashed, dotted) lines show the $1\sigma$ ($2\sigma$, $3\sigma$) contours.}
    \label{fig:appb_ppd_fbeB}
\end{figure}

A similar fit was performed for the magnetic energy parameter $\epsilon_B$. In this scenario, we once again fixed $n_0=1$ cm$^{-3}$ and performed a scan for $f_b\eta$ and $\epsilon_B$. The resulting posterior probability density is shown in Fig.~\ref{fig:appb_ppd_fbeB}. We see from this parameter scan that the variables are quite degenerate, so choosing a different value for $\epsilon_B$ than the one used in the main analysis will significantly change the constraints on $f_b$.

\end{appendix}
\end{document}